\documentclass[12pt]{elsart}
\usepackage{amsmath}
\usepackage{amssymb}
\usepackage{rotating}
\usepackage{color}
\usepackage{graphicx}
\usepackage{natbib}

\usepackage{url}

\usepackage{epstopdf}
\newcommand{\bm}[1]{\mathbf{#1}}
\newcommand{\bs}[1]{\boldsymbol{#1}}

\newcommand{\pabl}[2]{\frac{\partial #1}{\partial #2}}
\newcommand{\jump}[1]{\left[| #1 |\right]}
\newcommand{\abs}[1]{\left| #1 \right|}

\newcommand{\averg}[1]{\left\langle #1 \right\rangle}
\newcommand{\halb}{\frac{1}{2}}

\newcommand{\pl}{\,\,\,}

\newcommand{\dd}{\rm \, d}

\newcounter{eqnstop}

\begin{document}

\bibliographystyle{elsart-harv}

\renewcommand{\note}[1]{{\small \sl #1}}

\unitlength 1 cm

\begin{frontmatter}
\title{
Effective intrinsic linear properties of laminar piezoelectric composites and simple ferroelectric domain structures
}


\author{Johannes R\"odel}

\address{Technische Universit\"at Dresden\\
Institut f\"ur Werkstoffwissenschaft\\
Professur f\"ur Materialwissenschaft und Nanotechnik\\
Hallwachsstra{\ss}e 3\\
D-01062 Dresden, Germany\\[1ex]
}
\ead{roede@tmfs.mpgfk.tu-dresden.de	l}
\ead[url]{\url{http://www.mpgfk.tu-dresden.de/~roedel}}

\begin{abstract}\label{abstract}
% Text of abstract

The effective properties of piezoelectric laminates have been analyzed, based on the calculation of internal fields and making use of a simple matrix manipulation method. The results are expressed in a compact notation which is convenient for numerical implementation and at the same time 
suitable for further analytical treatments.
A detailed analysis of fully compatible ferroelectric domain structures shows, that the results for arbitrary piezoelectric laminates can be further simplified and specific property relationships for rank-1 laminates of tetragonal and rhombohedral crystals are derived.  
The method is finally applied to the analysis of various hierarchical domain structures. Detailed orientation relationships between the particular domains in some important domain pattern are given to make these structures accessible for the presented method.
Some numerical results for tetragonal barium titanate illustrate the effects of different domain arrangements on the effective properties.

\end{abstract}


\end{frontmatter}

%\url{http://www.mpgfk.tu-dresden.de/~roedel}

%Table of Contents
\newpage
\tableofcontents
\newpage

\newpage

\section{Introduction}

Modern functional materials, e.g. ferroelectrics, are often characterized by an internal microstructure, which in many cases can be described as a laminar structure. 
Although this is usually a simplification which neglects some effects, it is a useful approach that allows an analytical analysis of the overall properties of these materials. It provides insight into the physical properties of the composite, which may strongly differ from the local, single-phase properties.

In particular, we are interested in the linear material properties of a laminar compound and consider elastic, dielectric and piezoelectric coefficients as well as spontaneous strain and polarization.
The general approach may be applied also to other coupled problems and associated linear properties, if the relevant boundary conditions are appropriately chosen. In this context, magnetoelectric laminar composites \citep[e.g.]{avellaneda.94,nan.05} should be mentioned

The relation between microstructure, local and effective properties is one of the central subjects of the theory of composites \citep{milton.02}. 
Laminates or stratified materials are the simplest conceivable composites with a variation of the material properties in only one direction, the direction of lamination $\bs n$ \citep{milton.02}. These simple type of composites are also called rank-one laminates \citep{milton.02} or 2-2 composites \citep{newnham.78}.
For an overview of the historical development of the lamination theory and general solution strategies for various coupled problems see, for instance, \citet{milton.02}.

% Literatur Laminate: 
% Maxwell, JC Treatise ... (1873), pp.317-372 Oxford: laminates of laminates, eff. conductivity
% Bruggemann: Thesis, Utrecht, 1930. elastic moduli- stimmt das, steht das auch im Artikel?
% Backus (1962), J. Geophys. Res. 67, p. 4427-4440, allgemeine Loesung fuer Felder und Moduli

\begin{figure}[hbtp]
\begin{center}
\begin{picture}(10,8)
%\put(0,0){\includegraphics[width=10cm]{figures/PZT_T_AA.png}}
\put(0,0){\includegraphics[width=10cm]{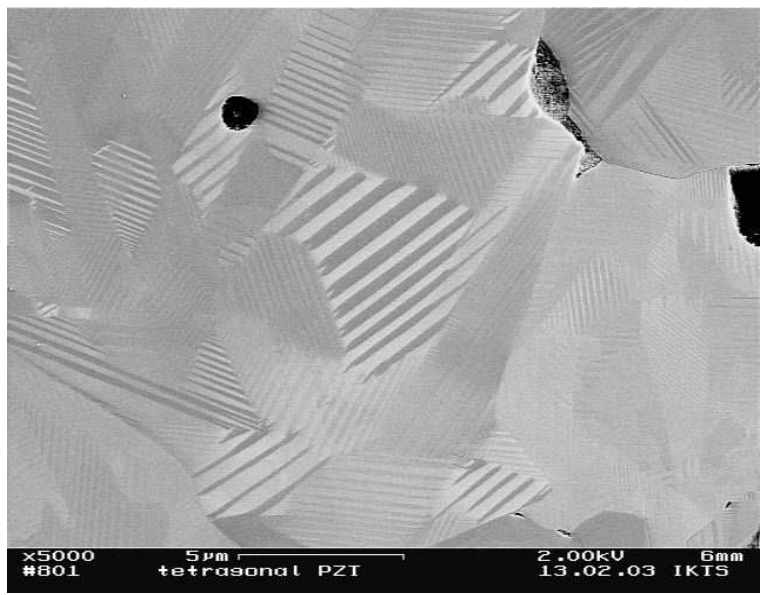}}
%\put(0,0){\framebox(10,7.8){}}
\end{picture}
\caption{\label{SEMdomains} SEM image of a ferroelectric ceramic (tetragonal PZT), showing a domain structure, which consists of differently oriented blocks of laminar domain patterns. }
\end{center}
\end{figure}

The knowledge of the effective behavior of piezoelectric multilayers is a crucial problem not only for 
stratified piezoelectric materials, which have been investigated for various transducer applications
\citep{cao.93b,zhang.94b,janas.95,lous.00,koh.05}.
A special case of a laminar piezoelectric structure is a ferroelectric crystal with a laminar domain structure, i.e.  a ferroelectric solid, which consists of two different ferroelectric
domain types (same crystal structure in the two phases, but with different crystal orientations with respect to the interface) separated by a crystallographically defined interface. 
The domains are arranged in laminar configuration and separated by plane 
interfaces (domain walls). An example of such a structure is shown in fig. \ref{SEMdomains}. The picture also demonstrates, that real microstructures are often much more complex. However, laminar stuctures can be used as building blocks in the description of more realistic patterns by assembling differently oriented laminates into a complex, often hierarchical microstructure of a single grain or crystallite.

In a recent review, \cite{bhattacharya.03} have pointed out the importance of the knowledge of the overall behavior of ferroelectric domain patterns for large strain piezoelectrics.
The problem of laminar ferroelectric domain structures becomes also an important research object in the context of engineered domain structures \citep{park.99,wada.99,zhang.01,yin.02,liu-li.03,liu-li.04,li.04,davis.05,wada.05}.

The purpose of this contribution is to provide a self-contained presentation of the theory of piezoelectric laminates, ranging from arbitrary stratified and electromechanically coupled multi-layers
to simple as well as hierarchical ferroelectric domain patterns. The theory offers both effective material properties (elastic, piezoelectric and dielectric constants) and average spontaneous strains and polarizations as well as internal fields within the layers and allows an analysis of effective (or averaged) symmetry properties of a multi-layer or multi-domain crystal. We do not restrict ourself to certain crystal symmetries or volume fractions.

Various variants of the problem and solutions have been published in the past \citep{turik.70engl,turik.74b,akcakaya.88,report.MW282710,gibiansky.99,erhart.99,erhart.01,bednarcyk.03,liu-li.03,li.04}. 
Although the solution of internal fields in piezoelectric multi-layers and the calculation of effective properties is straightforward and can be analytically derived without any difficulty \citep{milton.02}, this problem is worth to be revised again. In particular, because the analytical expressions are generally quite intricate, any contribution which provides a simplified approach should be welcome.
Additionally, the various authors concentrate on various but separate aspects of the problem and not all published results are correct.

The author has used the presented models in previous works on tetragonal ferroelectric ceramics \citep{roedel.03}.  Here, we generalize the approach to cover a variety of cases, ranging from arbitrary piezoelectric multilayers to particular ferroelectric domain structures with different crystal symmetries.

After stating the problem in section \ref{formulation}, it is therefore useful to review in section \ref{arbitrary_laminate}  a solution for the general problem of a piezoelectric multilayer with arbitrary number and structure of the constituent layers, which may apply for arbitrary piezoelectrically coupled laminates. 
The aim of our approach is to notate the internal fields and effective properties in a compact form, which is, at the same time, easy to implement numerically and convenient for further analytical treatments.

In section \ref{compatible_domains} we will consider the degenerate case of a fully compatible ferroelectric domain structure where certain orientation relationships are met. Further, we will give some examples for special cases with particular crystal structures. Then, domain patterns with incompatible polarizations are analyzed in contrast to the fully compatible domain structures.
Finally, we will briefly consider hierarchical laminates, in particular some important cases of hierarchical domain structures.

In the case of ferroelectric domain structures, one has to note, that domain walls are generally mobile and can move under applied fields. This may cause additional non-linear contributions to the macroscopic behavior, which is beyond the scope of this paper. However, the proposed approach provides also a convenient basis to describe non-linear reversible and irreversible response due to reversible or irreversible domain wall motion.

 \newpage

\section{Problem formulation and notation}\label{formulation}

\begin{figure}[hbtp]
\unitlength 0.875 cm % 16 cm entspricht 100%
\begin{center}
\begin{picture}(16,10)
%\put(0,0){\framebox(16,10){}}
%\put(0,0){\includegraphics[width=16cm]{figures/homogenlaminate.eps}}
\put(0,0){\includegraphics[width=14cm]{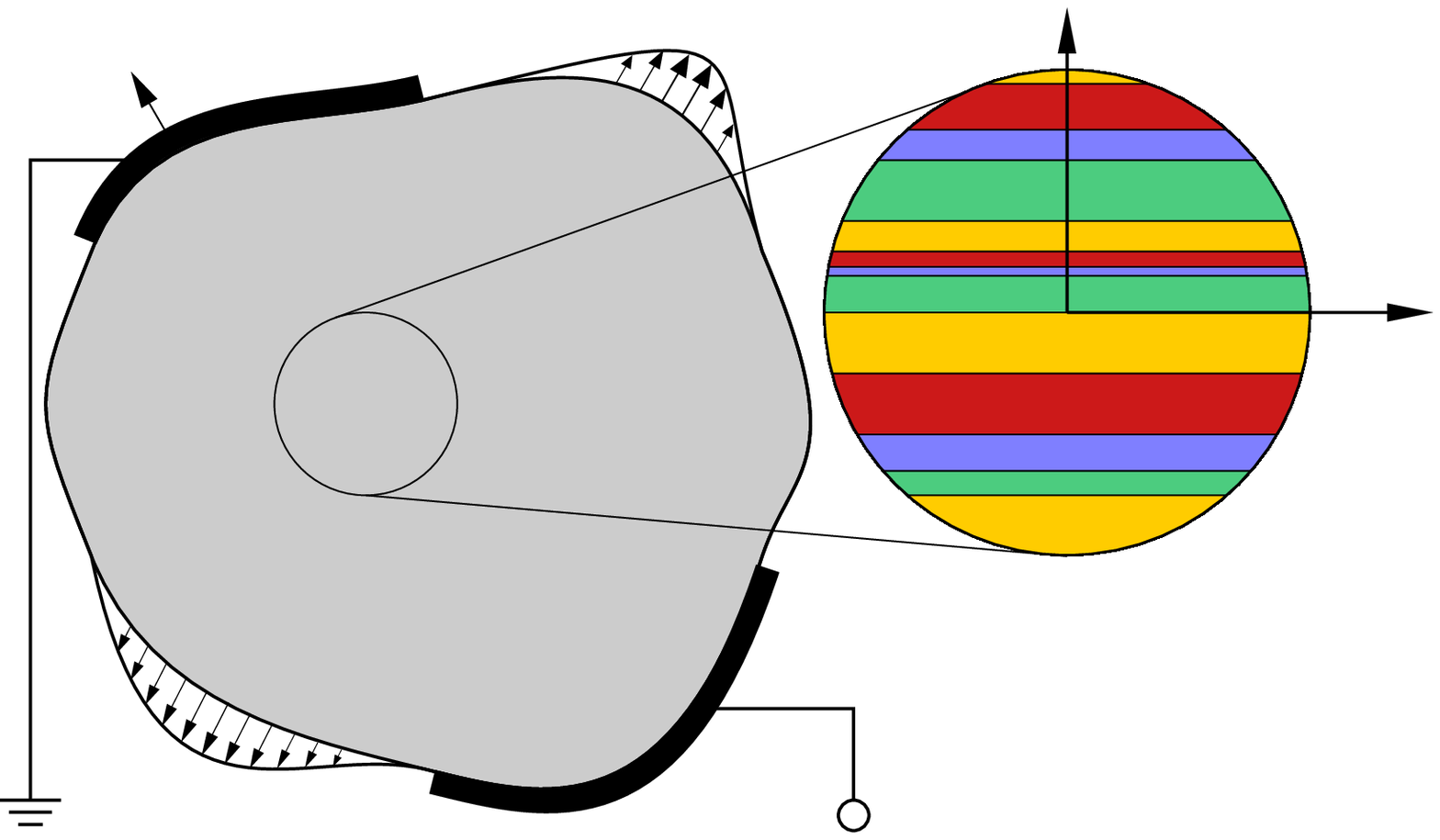}}
\put(9.8,0.5){\makebox(0,0){\Large $\phi$}}
\put(15.5,5.5){\makebox(0,0){\Large $x_1$}}
\put(11,9.2){\makebox(0,0){\Large $x_3$}}
\put(1.,8.5){\makebox(0,0){\Large $\bs n$}}
\put(7.3,9.3){\makebox(0,0){\Large $\bs t$}}
\put(1.7,1){\makebox(0,0){\Large $\bs t$}}
\end{picture}
\caption{\label{twsolid} Piezoelectric body subjected to prescribed tractions
and electric potentials. The body has a laminar micro-structure. }
\end{center}
\end{figure}

We consider a piezoelectric solid (see fig. \ref{twsolid}) under so-called homogeneous (and quasistatic) boundary conditions, i.e. boundary condition at the outer boundary of the solid are chosen so that internal fields of stress, $\bs \sigma(\bs x)$, strain $\bs \gamma(\bs x)$, electric field $\bs E(\bs x)$ and dielectric displacement $\bs D(\bs x)$ would be homogeneous, if the material properties would be homogeneous.

The electric fields and the linear elastic strain are defined by the gradients of the electric potential $\phi(\bs x)$ and elastic displacements $\bs u(\bs x)$, respectively:
\begin{equation}\label{kompatibel}
\gamma_{\rm ij}(\bs x) = \halb \left[ \pabl{}{x_{\rm j}} u_{\rm i}(\bs x) +\pabl{}{x_{\rm i}} u_{\rm j}(\bs x)\right], \qquad
E_{\rm i} (\bs x) =  - \pabl{}{x_{\rm i}} \phi (\bs x) 
\end{equation}

Surface tractions, $\bs t$, may be applied at the outer boundary with normal vector $\bs n$. 
The tractions are related to the elastic stress stress tensor $\bs \sigma (\bs x)$
\begin{equation}\label{OFkraft}
t_{\rm i} (\bs x) = \sigma_{\rm ij} (\bs x)\, n_{\rm j} (\bs x)
\end{equation}

Stress fields and electric fields have to fulfill the equilibrium conditions
\begin{equation}\label{equilibrium}
\pabl{}{x_{\rm j}} \sigma_{\rm ij} (\bs x)  = 0, \qquad
\pabl{}{x_{\rm i}} D_{\rm i} (\bs x)  = 0,
\end{equation}
where we neglect volume forces and free charges.
The equilibrium conditions imply the following equation at internal interfaces:
\begin{equation}\label{}
n_{\rm i} \jump{\sigma_{\rm ij}} = 0, \qquad 
n_{\rm i} \jump{D_{\rm i}} = 0,
\end{equation}
where $\jump{X}$ denotes the jump of the quantity $X$ at the interface and 
$\bs n$ is the normal vector of the interface plane.

We define the macroscopic fields by the volume averages over the local fields:
\begin{eqnarray}
\bar{\bs \sigma} = \frac{1}{V} \int\limits_V \bs \sigma (\bs x) \dd \bs x = \averg{\bs \sigma (\bs x)},&\qquad&
\bar{\bs \gamma} = \frac{1}{V} \int\limits_V \bs \gamma (\bs x) \dd \bs x = \averg{\bs \gamma (\bs x)}\nonumber \\[0.5ex]
\bar{\bs E} = \frac{1}{V} \int\limits_V \bs E (\bs x) \dd \bs x = \averg{\bs E (\bs x)},&\qquad&
\bar{\bs D} = \frac{1}{V} \int\limits_V \bs D (\bs x) \dd \bs x = \averg{\bs D (\bs x)},
\label{micro-macro}
\end{eqnarray}
which are the so-called micro-macro-relations.

The local fields are related to each other by the linear piezoelectric constitutive law:
{\setcounter{eqnstop}{\arabic{equation}}
\addtocounter{eqnstop}{1} % bisher steht da der letzte Wert drin!
\setcounter{equation}{0}
\renewcommand{\theequation}{\arabic{eqnstop}.\alph{equation}}
\begin{eqnarray}
\gamma_{\rm ij} &=& S_{\rm ijkl}^E\, \sigma_{\rm kl} \: 
+ \: d_{\rm ijk}\, E_{\rm k} \:  + \: %\left.
\gamma_{\rm ij}^{\rm s} %\right|_{E}
\label{eq_gamma}
\\[0.5ex]
D_{\rm i} &=& d_{\rm ikl}\, \sigma_{\rm kl} \: + \: \varepsilon_{\rm ik}^\sigma\, E_{\rm k} \: 
+ \: %\left. 
P_{\rm i}^{\rm s} %\right|_{\sigma},
\label{eq_D}
\end{eqnarray}
\setcounter{equation}{\arabic{eqnstop}}
} 
where  $S_{\rm ijkl}^E$ is the elastic stiffness tensor under constant electric fields,
$\varepsilon_{\rm ik}^\sigma$ is the dielectric permittivity tensor under constant stress and
$d_{\rm ijk}$ is the tensor of the piezoelectric strain constants, relating the strain to the electric field.

Ferroelectric crystals exhibit a spontaneous strain and spontaneous polarization,
$\bs \gamma^{\rm s}$ und $\bs P^{\rm s}$ respectively, at temperatures below the curie temperature.
The spontaneous strain is sometimes called stress free strain or eigenstrain, because it is caused by the ferroelectric phase transformation only, and not by any applied stress.
The reference state is a cubic state.
Correspondingly, the spontaneous polarization is the polarization of a crystal without applied electric fields. 
In the case, that thermoelastic and pyroelectric effects are of interest, the spontaneous strain and the spontaneous polarization can be extended using the thermal expansion and pyroelectric coefficients. 

One should be careful if other field quantities will be kept constant. For instance, the strain under constant dielectric displacement is not equal to the that one under constant electric field.

The constitutive equations can be conveniently expressed using Voigt's short index notation (reduced index notation). A compact notation of the electromechanical constitutive law, is obtained by combining equivalent mechanical and electrical field quantities in vectors with 9 components:
\begin{equation} 
{\bm p} = \left(\begin{array}{r} {\sigma}_\alpha \\ { E}_{\rm i}\end{array}
\right),
\quad
{\bm q} = \left(\begin{array}{r} { \gamma}_\alpha \\ { D}_{\rm i}\end{array}
\right),
\quad
{\bm q^{\rm s}} = \left(\begin{array}{r}
                              {\gamma}_\alpha^{\rm s}\\ 
                              {D}_{\rm i}^{\rm s}
                              \end{array}\right)
\qquad
\begin{array}{l}
\alpha = 1, 2,...6 \\
{\rm i} = 1,2,3
\end{array}
\end{equation}
\begin{equation} \label{local_const_law}
\bm q = \bm Q \, \bm p + \bm q^{\rm s}, \qquad 
{\bm Q} = \left(\begin{array}{r r} S_{\alpha,\beta}^E &  d_{\rm \alpha, j} \\ 
d_{\rm i, \beta} &  \varepsilon_{\rm ij}^\sigma
\end{array}\right),
\qquad
\begin{array}{l}
\alpha,\beta = 1, 2,...6 \\
{\rm i,j} = 1,2,3
\end{array}
\end{equation} 
and the inverted constitutive law is
\begin{equation} \label{inv_local_const_law}
\bm p = \bm R \, \left(  \bm q- \bm q^{\rm s}  \right), \qquad 
{\bm R} = {\bm Q}^{-1} = \left(\begin{array}{r r} C_{\alpha,\beta}^D &  -h_{\rm \alpha, j} \\ 
-h_{\rm i, \beta} &  \beta{\rm ij}^\gamma
\end{array}\right),
\qquad
\begin{array}{l}
\alpha,\beta = 1, 2,...6 \\
{\rm i,j} = 1,2,3
\end{array}
\end{equation} 

Combined property matrices, $\bm Q$ and $\bm R$ are symmetric.
Here, we have combined strain and dielectric displacement, as well as stress and electric field, because they are thermodynamically equivalent and  we will see later, that some results will be significantly simplified, if stress and electric fields are prescribed. However, it should be noted that mechanical stress and dielectric displacement as well as strain and negative electric field are equivalent from mathematical point of view:
\begin{equation} 
{\bm j} = \left(\begin{array}{r} {\sigma}_\alpha \\ { D}_{\rm i}\end{array}
\right),
\quad
{\bm f} = \left(\begin{array}{r} { \gamma}_\alpha \\ -{ E}_{\rm i}\end{array}
\right),
\quad
{\bm f^{\rm s}} = \left(\begin{array}{r}
\left.{\gamma}_\alpha^{\rm s}\right|_{D}\\ 
\left.-{E}_{\rm i}^{\rm s}\right|_{\sigma}
\end{array}\right),
\end{equation}
\begin{equation} 
\bm j  = \bm L \, \bm f + \bm j^{\rm s},
\qquad
{\bm L} = \left(\begin{array}{r r} {C}_{\alpha\beta}^E & {e}_{\alpha \rm j}^t \\ 
{e}_{\rm i \beta} & -{\varepsilon}_{\rm ij}^\gamma\end{array}\right),
\end{equation}

The above introduced combination of field quantities is applied to local fields as well as to the macroscopic, averaged fields.
The intention of this work is to calculate the effective material properties which fulfill the following equation describing the macroscopic linear material behavior:
\begin{equation}\label{avrg_const_law}
\bm{\bar q} = \bm Q^\ast \, \bm{\bar p} + {\bm q^\ast}^{\rm s},
\end{equation}
with 
\[
{\bm {\bar p}} = \left(\begin{array}{r} \bs {\bar \sigma} \\ \bs {\bar E} \end{array}\right) 
= \averg{\bm p(\bs x)} = \left(\begin{array}{r} \averg{\bs \sigma (\bs x)} \\ \averg{\bs  E (\bs x)} \end{array}\right) ,
\qquad
{\bm {\bar q}} = \left(\begin{array}{r} \bs {\bar \gamma} \\ \bs {\bar D} \end{array}\right) 
= \averg{\bm q(\bs x)} = \left(\begin{array}{r} \averg{\bs \gamma(\bs x)} \\ \averg{\bs  D(\bs x)} \end{array}\right)
\]
\[ 
{\bm q^{\rm \ast s}} = \left(\begin{array}{r}
                              {\bs \gamma}^{\ast \rm s}\\ 
                              {\bs P}^{\ast \rm s}
                              \end{array}\right), \qquad 
{\bm Q} = \left(\begin{array}{c c} \bs S^{\ast E} & \bs  d^{\ast t} \\ 
\bs d^\ast  & \bs  \varepsilon^{\ast \sigma}
\end{array}\right),
\] 

Following an often used ansatz in the homogenization theory we will express the local fields by disturbance of the external field
\begin{equation} \label{local_field_p}
\bm p = \bm A \bm {\bar p} + \bm p^A,
\end{equation}
where $\bm A$ is an interaction matrix, which describes the interaction of the various phases under applied loads $\bm {\bar p} $ and $\bm p^A$ is a vector which describes stresses and electric fields, which arise in the lamella $(k)$ due to incompatible spontaneous stresses and incompatible spontaneous polarizations.

The general formalism to calculate the effective material properties is to include \eqref{local_const_law} and \eqref{local_field_p} into \eqref{avrg_const_law}
\begin{equation}
\bm {\bar q} = \averg{\bm q} = \averg{\bm Q  \bm p + \bm q^s} 
= \averg{\bm Q  (\bm A \bm {\bar p} + \bm p^A ) + \bm q^s} 
= \averg{\bm Q  \bm A} \bm {\bar p} + \averg{\bm Q \bm p^A  + \bm q^s},
\end{equation}
from which we derive the general expression for the effective properties by comparison with equation \eqref{avrg_const_law}:
\begin{equation}\label{effectiv_prop}
 \bm Q^\ast = \averg{\bm Q  \bm A} , \qquad \bm q^{\ast \rm s} =  \averg{\bm Q \bm p^A  + \bm q^s}
\end{equation}

The remaining task is to calculate the coefficients of the matrix $\bm A$ and the vector $ \bm p^A $

\newpage

\section{Arbitrary piezoelectric laminar structures}\label{arbitrary_laminate}

In order to derive a solution for a piezoelectric multilayer further assumptions are necessary.
We consider a piezoelectric solid, which has a layered microstructure consisting of $n$ individual layers.
Each layer $k$ may have different electromechanical properties:
\[
\bm Q^{(k)}=
\left(
\begin{array}{ccc ccc ccc}
S_{11}^{(k)}  &  S_{12}^{(k)}  & S_{13}^{(k)}  & S_{14}^{(k)} & S_{15}^{(k)} & S_{16}^{(k)} & d_{11}^{(k)}  & d_{21}^{(k)} &  d_{31}^{(k)} \\
S_{12}^{(k)}  &  S_{22}^{(k)}  & S_{23}^{(k)}  & S_{24}^{(k)} & S_{25}^{(k)} & S_{26}^{(k)} & d_{12}^{(k)}  & d_{22}^{(k)} &  d_{32}^{(k)} \\
S_{13}^{(k)}  &  S_{23}^{(k)}  & S_{33}^{(k)}  & S_{34}^{(k)} & S_{35}^{(k)} & S_{36}^{(k)} & d_{13}^{(k)}  & d_{23}^{(k)} &  d_{33}^{(k)} \\
S_{14}^{(k)}  &  S_{24}^{(k)}  & S_{34}^{(k)}  & S_{44}^{(k)} & S_{45}^{(k)} & S_{46}^{(k)} & d_{14}^{(k)}  & d_{24}^{(k)} &  d_{34}^{(k)} \\
S_{15}^{(k)}  &  S_{25}^{(k)}  & S_{35}^{(k)}  & S_{45}^{(k)} & S_{55}^{(k)} & S_{56}^{(k)} & d_{15}^{(k)}  & d_{25}^{(k)} &  d_{35}^{(k)} \\
S_{16}^{(k)}  &  S_{26}^{(k)}  & S_{36}^{(k)}  & S_{46}^{(k)} & S_{56}^{(k)} & S_{66}^{(k)} & d_{16}^{(k)}  & d_{26}^{(k)} &  d_{36}^{(k)} \\
d_{11}^{(k)}  &  d_{12}^{(k)}  & d_{13}^{(k)}  & d_{14}^{(k)} & d_{15}^{(k)} & d_{16}^{(k)} & \varepsilon_{11}^{(k)}  & \varepsilon_{12}^{(k)} &  \varepsilon_{13}^{(k)} \\
d_{21}^{(k)}  &  d_{22}^{(k)}  & d_{23}^{(k)}  & d_{24}^{(k)} & d_{25}^{(k)} & d_{26}^{(k)} & \varepsilon_{12}^{(k)}  & \varepsilon_{22}^{(k)} &  \varepsilon_{23}^{(k)} \\
d_{31}^{(k)}  &  d_{32}^{(k)}  & d_{33}^{(k)}  & d_{34}^{(k)} & d_{35}^{(k)} & d_{36}^{(k)} & \varepsilon_{13}^{(k)}  & \varepsilon_{23}^{(k)} &  \varepsilon_{33}^{(k)} 
\end{array}
\right), \qquad
\bm q^s = \left(
\begin{array}{c}
{\gamma_1^s}^{(k)}\\
{\gamma_2^s}^{(k)}\\
{\gamma_3^s}^{(k)}\\
{\gamma_4^s}^{(k)}\\
{\gamma_5^s}^{(k)}\\
{\gamma_6^s}^{(k)}\\
{P_1^s}^{(k)}\\
{P_2^s}^{(k)}\\
{P_3^s}^{(k)}
\end{array}
\right)
\]
Inside each layer, the material properties are constant.

The interfaces between the layers are considered to be flat, plane and parallel to each other.
As the direction of lamination we have chosen the $x_3$-direction, i.e. the macroscopical  $x_3$-direction should be normal to the interfaces.
Additionally, we assume that every layer is very thin compared to its length. 
Then, we can neglect effects at the edges and the volume fraction of of the $k$th layer, $\xi^k$, is proportional to it's thickness.

Under this assumptions, we can presume that the internal fields do not vary in the directions perpendicular to direction of lamination, i.e. the internal fields depend only on $x_3$.
Then, the equilibrium equations imply that the following field components are constant within the layers and continuous at the boundary between two layers i and j, which means, that they are equal to the applied fields: 
{\setcounter{eqnstop}{\arabic{equation}}
\addtocounter{eqnstop}{1} % bisher steht da der letzte Wert drin!
\setcounter{equation}{0}
\renewcommand{\theequation}{\arabic{eqnstop}.\alph{equation}}
\begin{alignat}{2}
\label{stackbed1}
\sigma_{33}^{\rm i} &= \sigma_{33}^{\rm j} = \bar{\sigma}_{33} \pl &\longrightarrow \pl p_3^{(k)} &= \bar p_3 \\ 
\label{stackbed2}
\sigma_{13}^{\rm i} &= \sigma_{13}^{\rm j} = \bar{\sigma}_{13} \pl &\longrightarrow \pl p_5^{(k)} &= \bar p_5 \\ 
\sigma_{23}^{\rm i} &= \sigma_{23}^{\rm j} = \bar{\sigma}_{23} \pl &\longrightarrow  \pl p_4^{(k)} &= \bar p_4 \label{stackbed3}\\ 
D_{3}^{\rm I} &= D_{3}^{\rm II} %= \bar{D}_{3} 
\pl &\longrightarrow \pl q_9^{(k)} &=  q_9^{(1)} %\bar q_9  
\label{stackbed4}\\
\intertext{
Since the electric field has to be curl free and strains have to be compatible, the following fields components are continuous across the interfaces and following conditions must be satisfied:}
\gamma_{11}^{\rm i} &= \gamma_{11}^{\rm j} %= \bar{\gamma}_{11} 
\pl &\longrightarrow \pl q_1^{(k)} &= q_1^{(1)}
%\bar q_1 
\label{stackbed5}\\ 
\gamma_{22}^{\rm i} &= \gamma_{22}^{\rm j} %= \bar{\gamma}_{22} 
\pl &\longrightarrow \pl q_2^{(k)} &= q_2^{(1)}%\bar q_2 
\label{stackbed6}\\ 
\gamma_{12}^{\rm i} &= \gamma_{12}^{\rm j} %= \bar{\gamma}_{12} 
\pl &\longrightarrow \pl q_6^{(k)} &= q_6^{(1)}%\bar q_6 
\label{stackbed7}\\ 
E_{1}^{\rm i} &= E_{1}^{\rm j} = \bar{E}_{1} \pl &\longrightarrow \pl p_7^{(k)} &= \bar p_7  \label{stackbed8}\\
E_{2}^{\rm i} &= E_{2}^{\rm j} = \bar{E}_{2} \pl &\longrightarrow \pl p_8^{(k)} &= \bar p_8  \label{stackbed9}
\end{alignat}
\setcounter{equation}{\arabic{eqnstop}}
} 
Additionally, within a  particular layer, all fields are constant. As a result, the volume average of a property $X$ is the sum of $X^{(k)}$, weighted  with the volume fraction $\xi^{(k)}$
\[
\averg{X} = \sum_k \xi^{(k)} X^{(k)}
\]

We summarize the conditions, which have to be satisfied:
{\setcounter{eqnstop}{\arabic{equation}}
\addtocounter{eqnstop}{1} % bisher steht da der letzte Wert drin!
\setcounter{equation}{0}
\renewcommand{\theequation}{\arabic{eqnstop}.\alph{equation}}
\begin{alignat}{2}
q_{\rm i}^{(k)} &= q_{\rm i}^{(1)}%\bar q_{\rm i} 
\qquad &{\rm i} &= 1,2,6,9 \label{stackcond_q}
\\
p_{\rm i}^{(k)} &= \bar p_{\rm i} & {\rm i} &= 3,4,5,7,8 \label{stackcond_p1}\\
\intertext{As we want solve a set of equations \eqref{local_field_p} for the internal fields $\bm p$, we need four additional equations. We make use of the micro-macro-relation \eqref{micro-macro}:}
\averg{p_{\rm i}^{(k)}} &= \bar p_{\rm i} & {\rm i}&= 1,2,6,9 \label{stackcond_p2}
\end{alignat}
\setcounter{equation}{\arabic{eqnstop}}
}

It is furthermore convenient to split the indices into two groups, according to which boundary condition holds for $p_i$:
\[
\begin{array}{ll}
{\rm \tilde i} &= 1,2,6,9 \\
{\rm \tilde{\tilde i}} &= 3,4,5,7,8
\end{array}
\]
The calculation of the components of $\bm A$ and $\bm p^A$ is straightforward and given in detail in the appendix.
We summarize the solution in the following form:
{\setcounter{eqnstop}{\arabic{equation}}
\addtocounter{eqnstop}{1} % bisher steht da der letzte Wert drin!
\setcounter{equation}{0}
\renewcommand{\theequation}{\arabic{eqnstop}.\alph{equation}}
\begin{alignat}{3}
A_{\rm ik}^{\rm Q\, (p)} &= \delta_{\rm ik}      					&\rm i&={\rm \tilde{\tilde i}} \quad& \rm k&=1..9
\label{sum_arbsol_a} \\[1ex]
A_{\rm ik}^{\rm Q\,(p)} &= R_{\rm i\tilde m}^{\rm (p)}  \widehat{Q}_{\rm \tilde m k} &\rm i&={\rm \tilde i} &\rm k&={\rm \tilde k}
\label{sum_arbsol_b} \\[1ex]
A_{\rm ik}^{\rm Q\,(p)} &= 
\tilde{R}_{\rm \tilde i\tilde m}^{\rm (p)} \widehat{Q}_{\rm \tilde m \tilde n}
\averg{ \tilde{R}_{\rm \tilde n \tilde p}^{\rm (p)} Q_{\rm \tilde p k}^{\rm (p)} }
- \tilde{R}_{\rm \tilde i\tilde m}^{\rm (p)} Q_{\rm \tilde m k}^{\rm (p)}  \qquad\qquad&\rm \tilde i&={\rm \tilde i} &\rm k&={\rm \tilde{\tilde k}} 
\label{sum_arbsol_c} \\[2ex]
p_{\rm i}^{\rm A\, (p)} &= 0 						&&	&\rm i &= {\rm \tilde{\tilde i}}  \label{sum_arbsol_d} \\[1ex]
p_{\rm i}^{\rm A\, (p)} &= 
\tilde{R}_{\rm \tilde i\tilde m}^{\rm (p)}  \widehat{Q}_{\rm \tilde m \tilde n} 
\averg{ \tilde{R}_{\rm \tilde n \tilde p}^{\rm (p)} q_{\rm \tilde p}^{\rm s\, (p)} } 
- \tilde{R}_{\rm \tilde i\tilde p}^{\rm (p)} q_{\rm \tilde p}^{\rm s\,(p)} 	&&	&\rm i &= {\rm \tilde i}
\label{sum_arbsol_e} 
\end{alignat}
\setcounter{equation}{\arabic{eqnstop}}
} 
where we made use of the following abbreviations: $\bm {\tilde Q}$ is a $4 \times 4$ submatrix of the property matrix $\bm Q$, and $\bm {\tilde R}$ is the inverse of  $\bm {\tilde Q}$.
\[
\tilde{R}_{\rm \tilde i \tilde k}^{\rm (p)} = \left(
\begin{array}{ccccc}
Q_{11}^{\rm (p)}  & Q_{12}^{\rm (p)}  & Q_{16}^{\rm (p)}  & Q_{19}^{\rm (p)}  \\
Q_{21}^{\rm (p)}  & Q_{22}^{\rm (p)}  & Q_{26}^{\rm (p)}  & Q_{29}^{\rm (p)}  \\
Q_{61}^{\rm (p)}  & Q_{62}^{\rm (p)}  & Q_{66}^{\rm (p)}  & Q_{69}^{\rm (p)}  \\
Q_{91}^{\rm (p)}  & Q_{92}^{\rm (p)}  & Q_{96}^{\rm (p)}  & Q_{99}^{\rm (p)}  
\end{array}\right)^{-1},
\]
and $\bm {\widehat{Q}}$ is the inverse matrix of the volume average of $\bm {\tilde R}$:
\[
{\bm {\widehat{Q}}} =  \averg{\bm {\tilde R}}^{-1} \quad
\mbox{or in index notation}\quad
\widehat{Q}_{\rm \tilde i \tilde k} =
\averg{\tilde{R}_{\rm\tilde i \tilde k}^{\rm (p)}}^{-1}
\]
Note that $\tilde Q_{\rm \tilde i \tilde k}$ is a $4\times 4$ sub-matrix   
of the $9 \times 9$ matrix $Q_{\rm ik}$. 
$\tilde{R}_{\rm \tilde i \tilde k}$ is the inverse of this $4 \times 4$ 
matrix  $\tilde Q_{\rm \tilde i \tilde k}$, and not a $4 \times 4$ sub-matrix of $\bm Q^{-1}$.
$\bm {\tilde R}$ is symmetric:
\[ 
\tilde{R}_{\rm \tilde i \tilde k} = \tilde R_{\rm \tilde k \tilde i} , \qquad
\tilde{R}_{\rm \tilde i \tilde j} Q_{\rm \tilde j \tilde k} = Q_{\rm \tilde i \tilde j} \tilde{R}_{\rm \tilde j \tilde k} =\delta_{\rm \tilde i \tilde k} 
\]

Then follows that $\bm A^{\rm (p)}$ and $\bm p^A$ have the form:
{\[
\setlength{\arraycolsep}{3pt}
\renewcommand{\arraystretch}{1.2}
A_{\rm ik}^{\rm (p)} = \left( 
\begin{array}{ccccccccc}
A_{11} & A_{12} & A_{13} & A_{14} & A_{15} & A_{16} & A_{17} & A_{18}  & A_{19} \\
A_{21} & A_{22} & A_{23} & A_{24} & A_{25} & A_{26} & A_{27} & A_{28}  & A_{29} \\
0 & 0 & 1 & 0      & 0      & 0 & 0      & 0      & 0 \\
0 & 0 & 0 & 1      & 0      & 0 & 0      & 0      & 0 \\
0 & 0 & 0 & 0      & 1      & 0 & 0      & 0      & 0 \\
A_{61} & A_{62} & A_{63} & A_{64} & A_{65} & A_{66} & A_{67} & A_{68} & A_{69} \\
0 & 0 & 0 & 0      & 0      & 0 & 1      & 0      & 0 \\
0 & 0 & 0 & 0      & 0      & 0 & 0      & 1      & 0 \\
 A_{91} &  A_{92} &  A_{93} &  A_{94} & A_{95} &  A_{96} & A_{97} &  A_{98} &  A_{99} 
\end{array}\right),
\qquad
p_{\rm i}^{\rm A\, (p)} = \left( 
\begin{array}{c}
p_1^{\rm A \, (p)} \\
p_2^{\rm A \, (p)} \\
0 \\
0 \\
0 \\
p_6^{\rm A \, (p)} \\
0 \\
0 \\
p_9^{\rm A \, (p)}
\end{array}\right)
\]
}
If we introduce the interaction matrices into \eqref{effectiv_prop} we find
{\setcounter{eqnstop}{\arabic{equation}}
\addtocounter{eqnstop}{1} % bisher steht da der letzte Wert drin!
\setcounter{equation}{0}
\renewcommand{\theequation}{\arabic{eqnstop}.\alph{equation}}
\begin{align}
Q_{\rm \tilde i \tilde k}^\ast 
&= \averg{Q_{\rm \tilde i \tilde j} A_{\rm \tilde j \tilde k} + Q_{\rm \tilde i \tilde{\tilde j}} A_{\rm \tilde{\tilde j} \tilde k}} 
= \widehat{Q}_{\rm \tilde i \tilde k}
\label{effprop_arb_a}\\[1ex]
Q_{\rm \tilde i \tilde{\tilde k}}^\ast 
&= \averg{Q_{\rm \tilde i \tilde j} A_{\rm \tilde j \tilde{\tilde k}} + Q_{\rm \tilde i \tilde{\tilde j}} A_{\rm \tilde{\tilde j} \tilde{\tilde k}}} 
= \averg{Q_{\rm \tilde i \tilde j} A_{\rm \tilde j \tilde{\tilde k}}} + \averg{Q_{\rm \tilde i \tilde{\tilde j}}} 
= \widehat{Q}_{\rm \tilde i \tilde j} \averg{\tilde{R}_{\rm \tilde j \tilde l} Q_{\rm \tilde l \tilde{\tilde k}} }
\label{effprop_arb_b}\\[1ex]
Q_{\rm \tilde{\tilde i} \tilde k}^\ast 
&= \averg{Q_{\rm \tilde{\tilde i} \tilde j} A_{\rm \tilde j \tilde k} + Q_{\rm \tilde{\tilde i} \tilde{\tilde j}} A_{\rm \tilde{\tilde j}\tilde k}} 
= \averg{ Q_{ \rm \tilde{\tilde i} \tilde j} \tilde{R}_{\rm \tilde j \tilde l} }  \widehat{Q}_{\rm \tilde l \tilde k} 
\label{effprop_arb_c}\\[1ex]
Q_{\rm \tilde{\tilde i} \tilde{\tilde k}}^\ast 
&= \averg{Q_{\rm \tilde{\tilde i} \tilde j} A_{\rm \tilde j \tilde{\tilde k}} + Q_{\rm \tilde{\tilde i} \tilde{\tilde j}} A_{\rm \tilde{\tilde j} \tilde{\tilde k}}} 
\nonumber \\
&= \averg{ Q_{ \rm \tilde{\tilde i} \tilde j} \tilde{R}_{\rm \tilde j \tilde l} }  \widehat{Q}_{\rm \tilde l \tilde m} 
\averg{\tilde{R}_{\rm \tilde m \tilde n} Q_{\rm \tilde n \tilde{\tilde k}} }
+\averg{ Q_{ \rm \tilde{\tilde i} \tilde j} \tilde{R}_{\rm \tilde j \tilde l}  Q_{\rm \tilde l \tilde{\tilde k}} }  
+\averg{Q_{\rm \tilde{\tilde i} \tilde{\tilde k} }}
\label{effprop_arb_d}\\[1ex]
q_{\rm \tilde i}^{\rm s \ast} & = \averg{Q_{\rm \tilde i \tilde j}  p_{\rm \tilde j}^A} + \averg{q_{\rm \tilde i}^{\rm s} }
=  \widehat{Q}_{\rm \tilde i \tilde j} \averg{  \tilde{R}_{\rm \tilde j \tilde k}  q_{\rm \tilde k}^{\rm s} } 
%+ \averg{q_{\tilde i}^{\rm s} }
\label{effprop_arb_e}\\[1ex]
q_{\rm \tilde{\tilde i}}^{\rm s \ast} & = \averg{Q_{ \rm \tilde{\tilde i} \tilde j}  p_{\rm \tilde j}^A} + \averg{q_{ \rm \tilde{\tilde i}}^{\rm s} }
\nonumber \\[1ex]
&= \averg{
Q_{ \rm \tilde{\tilde i} \tilde j} \tilde{R}_{\rm \tilde j \tilde k} } \widehat{Q}_{\rm \tilde k \tilde l} 
\averg{\tilde{R}_{\rm \tilde l \tilde m}  q_{\rm \tilde m}^{\rm s} }
+ \averg{
Q_{ \rm \tilde{\tilde i} \tilde j} \tilde{R}_{\rm \tilde j \tilde k}  q_{\rm \tilde k}^{\rm s} }
+ \averg{q_{\rm \tilde{\tilde i}}^{\rm s} }
\label{effprop_arb_f}
\end{align}
\setcounter{equation}{\arabic{eqnstop}}
} 
This equation for the effective properties are equivalent to those given by \citet{liu-li.03}, and \citet{li.04}.
Although their solution is very elegant, our derivation offers some advantages, as we have directly given the equations for the internal fields $\bm p^{(k)}$. Additionally, we have also derived the the relationship for the effective spontaneous strain and polarization of the composite.
In the often cited paper by \citet{erhart.99} the authors have used wrong conditions at the interfaces, which obviously led them to wrong conclusions.  The rather detailed treatment of the general case may be also seen as a correction of their results.

Alternative sets of effective properties, like $\bm L^\ast$ may be calculated with very similar formulas. The only difference is, that the separation of indices according to the boundary conditions, or likewise, the separation into sub-matrices as shown in the appendix, will be different \citep{report.MW282710}.

To close this section, we remind, that the effective properties have to obey the simple bounds (Voigt-Reuss-type bounds)
\begin{equation}
 \averg{\bm Q}^{-1} \le  \left( \bm Q^\ast \right)^{-1} = \bm R^\ast  \le \averg{\bm R} 
\end{equation}
or
\begin{equation}
\averg{\bm R}^{-1} \le \bm Q^\ast \le \averg{\bm Q} 
\end{equation}
where the inequalities are ment in the sense of positive definiteness.

\newpage

%%%%%%%%%%%%%%%%%%%%%%%%%%%%%%%%%%%%%%%%%%%%%%%%%%%%%%%%%%%%%%%%%%%%%%%%%%%%%%%%%%%%%%%%%%%%%%%%%%%%%%

\section{Simple compatible domain structures}\label{compatible_domains}\label{compatible-rank-1}

Ferroelectric domains are a result of symmetry breaking phase transformations. In the ferroelectric phase exist several equivalent variants with the same, but rotated, crystal structure.
Spontaneous polarization and spontaneous strain due to the phase transformation are strictly coupled to each other and correlated to the crystal lattice. The term {\em ferroelectric} refers to the fact, that the polarizations, and thus the strain, can be switched from one direction to an other by applied electric fields.

For sake of energy minimization, ferroelectric crystals usually exist in a multi-domain state and the interfaces between the domains are also correlated to the crystal lattice. This yields particular crystallographic orientation relationships between different domains.

Consider a material with arbitrary symmetry. 
On a microscopic scale (within the domains) the material is assumed to obey
linear piezoelectric behavior with the single crystal properties
$\bf Q$ and $\bf q^{\rm s}$. We do not consider nonlinear effects, neither switching nor any kind of domain wall motions. It should be emphasized that the present treatment can easily extended to take into account reversible and irreversible motions of the domain walls. This will be published elsewhere.

In this section we want to consider a fully compatible laminar arrangement of two domain types. The term {\em compatible} generally means, that the spontaneous strain and polarization of the two domains are compatible at the interface so that there appear neither stresses nor electric fields inside the domains, if there is no macroscopic stress and no macroscopic field. It is obvious, that this is the case, if the the components $\gamma_{\rm 11}^{\rm s}$, $\gamma_{\rm 12}^{\rm s}$ and $\gamma_{\rm 22}^{\rm s}$ of the spontaneous strain and $P_3^{\rm s}$ of the polarization are continuos across the interface, i.e. these components have to be homogeneous.
This requirement can also be demonstrated based on eqn \eqref{sum_arbsol_e}, which says that $\bm p = \bm 0$ if all $q_{\rm i}^s$ are homogeneous for i = 1,2,6,9; i.e. ${q_{\rm i}^s}^{ \rm I}= {q_{\rm i}^s}^{\rm II}; i = 1,2,6,9$.

\begin{figure}[hbtp]
\begin{center}
\begin{picture}(10,10)
\put(0,0){\includegraphics[width=10cm]{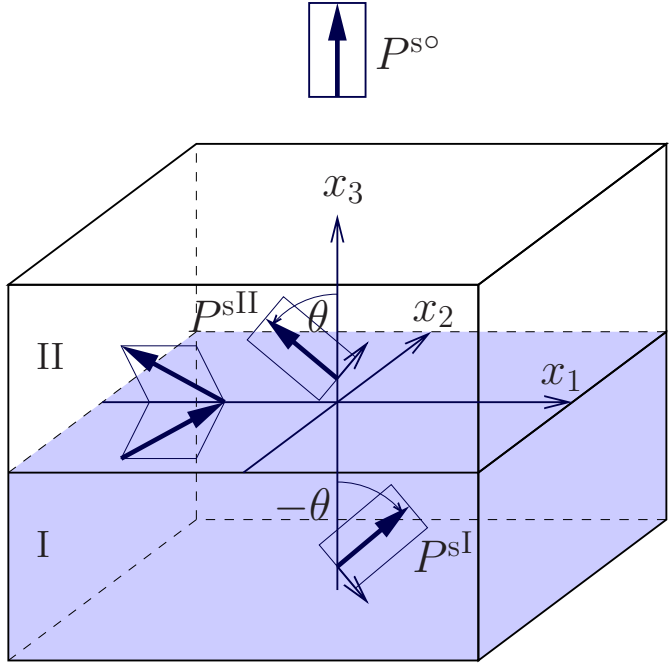}}
\put(7.7,4.3){\Large$ x_1$}
\put(6,5.15){\Large $x_2$}
\put(4.8,6.8){\Large$ x_3$}
\put(4.6,5.){\Large$\theta$}
\put(4.15,2.5){\Large $ -\theta$}
\put(1,2){\large I }
\put(1,4.5){\large II}
\put(5.5,8.5){\Large$ {P^{\rm s}}^\circ$}
\put(6,1.8){\Large$ {P^{\rm s}}^{\rm I}$}
\put(3,5){\Large$ {P^{\rm s}}^{\rm II}$}

%\put(0,0){\framebox(10,10){}}
\end{picture}
\caption{\label{domaenen_1} Orientation relationship of compatible ferroelectric  domains. Thick arrows represent polarizations. Two possible types of unit cells indicate spontaneous strains.}
\end{center}
\end{figure}

Assume, that we have a crystal in arbitrary orientation with respect to to the interface, with $\bs n = (0,0,1)^t $, on one side of the interface. 
For a particular crystal symmetry, only a limited set of orientations relationships between the crystal orientation and the interface (or domain wall) are possible. For a detailed analysis of the this relationships, including a refined compatibility condition which takes into account the change of crystal symmetry due to the ferroelectric phase transformation, further arguments are necessary \citep{shu.01}. However, this is beyond the scope of this paper.
We simply assume, that the plane, characterized by $\bs n = (0,0,1)^t $, is a possible domain interface for the particular crystal orientation.

A fully compatible domain can be created by a rotation of the crystal on the opposite side of the interface. 
The part of the crystal, which is in the original orientation, will be called domain I, denoted by the suffix I. likewise the rotated part will be domain  II.
The crystal coordinate system of domain II is then rotated by the following relationship:
\begin{equation}\label{orient-relation}
\bs x^{\rm II} = \bs \omega\, \bs x^{\rm I},
\end{equation}
where $\bs \omega$ is a rotation matrix

\begin{equation}
\label{rotation_comptbl}
\bs \omega = \left(
\begin{array}{rrr}
 -1 & 0  & 0  \\
  0 & \; -1 & \pl 0  \\
  0 & 0  & 1  
\end{array}
\right).
\end{equation}

It should be noted, that this rotation ensures only compatibility of spontaneous strain and polarization. It does not necessarily describe a pair of ferroelectric domains, which form as a result of a phase transformation.
This question should be considered separately for particular crystal symmetries.

The properties of the rotated crystal with respect to the global coordinate system are expressed by
\begin{alignat*}{3}
P_{\rm i}^{\rm II} &= \omega_{\rm ij} \, P_{\rm j}^{\rm I}, &\pl &\mbox{or} \pl  &\bs P^{\rm II} &= \bs \omega\, \bs P^{\rm I}
\\
\gamma_{\rm ij}^{\rm II} &=   \omega_{\rm ik}  \, \omega_{\rm jl}  \, \gamma_{\rm kl}^{\rm I} , &\pl &\mbox{or} \pl 
&\bs \gamma^{\rm II} &= \bs  \omega  \, \bs \gamma^{\rm I} \, \bs \omega^t 
\\
d_{\rm ijk}^{\rm II}  &= \omega_{\rm il}  \, \omega_{\rm jm} \, \omega_{\rm kn} \, d_{lmn}^{\rm I}
\\
S_{\rm ijkl}^{\rm II}  &= \omega_{\rm im} \,  \omega_{\rm jn} \, \omega_{\rm ko} \, \omega_{\rm lp} \, S_{mnop}^{\rm I}
\end{alignat*}

We find the following relationship for vectors, like polarizations:
\[
{\bs P^{\rm s}}^{\rm I} = \left(\begin{array}{r} P_1^{\rm s}  \\ P_2^{\rm s}  \\ P_3^{\rm s}    
\end{array}
\right), 
\quad 
{\bs P^{\rm s}}^{\rm II} = \left(\begin{array}{r} -P_1^{\rm s}  \\ -P_2^{\rm s}  \\ P_3^{\rm s}    
\end{array}
\right), \quad 
\]
For symmetric 2nd rank tensors like strain and permittivity holds
\[
\bs \gamma^{I} = \left( \begin{array}{rrr}
\gamma_{11} & \gamma_{12}   & \gamma_{13}   \\
& \gamma_{22}   & \gamma_{23}   \\
...& &  \gamma_{33}  
\end{array}\right), 
\quad
\bs \gamma^{II} = \left( \begin{array}{rrr}
\gamma_{11} & \pl \gamma_{12}   & \; -\gamma_{13}   \\
& \gamma_{22}   & -\gamma_{23}   \\
...& &  \gamma_{33}  
\end{array}\right),
\]
which shows, that the proposed  general orientation relationship between the two domains ensures compatibility of spontaneous strain and polarization, independent of the particular symmetry of the crystal.
For the piezoelectric coefficients and the elastic constant we get (in Voigt's notation):
\[
\bs d^{\rm I} = \left(
\begin{array}{rrr rrr}
d_{11} &  d_{12} & d_{13} & d_{14}& d_{15}& d_{16}\\
d_{21} &  d_{22} & d_{23} & d_{24}& d_{25}& d_{26}\\
d_{31} &  d_{32} & d_{33} & d_{34}& d_{35}& d_{36}
\end{array}\right),
\quad
\bs d^{\rm II} = \left(
\begin{array}{rrr rrr}
-d_{11} &  -d_{12} & -d_{13} & d_{14}& d_{15}& -d_{16}\\
-d_{21} &  -d_{22} & -d_{23} & d_{24}& d_{25}& -d_{26}\\
d_{31} &  d_{32} & d_{33} & -d_{34}& -d_{35}& d_{36}
\end{array}
\right)
\]

\[
\bs S^{\rm I}= \left( \begin{array}{rrr rrr }
S_{11} &  S_{12} & S_{13} & S_{14}& S_{15}& S_{16}\\
&  S_{22} & S_{23} & S_{24}& S_{25}& S_{26}\\
&  & S_{33} & S_{34}& S_{35}& S_{36}\\
&  ... & & S_{44}& S_{45}& S_{46}\\
&  & & & S_{55}& S_{56}\\
&  & & & & S_{66}
\end{array}\right),
\quad
\bs S^{\rm II}= \left( \begin{array}{rrr rrr }
S_{11} &  S_{12} & S_{13} & -S_{14}& -S_{15}& S_{16}\\
&  S_{22} & S_{23} & -S_{24}& -S_{25}& S_{26}\\
&  & S_{33} & -S_{34}& -S_{35}& S_{36}\\
&  ... & & S_{44}& S_{45}& -S_{46}\\
&  & & & S_{55}& -S_{56}\\
&  & & & & S_{66}
\end{array}\right)
\]

If we combine the properties of the domain II in the compact form we get
\[
\bm Q^{\rm II}=
\left(
\begin{array}{rrr rrr rrr}
S_{11} &  S_{12} & S_{13} & - S_{14}& -S_{15}& S_{16}& -d_{11} & -d_{21}&  d_{31}\\
& S_{22} & S_{23} & -S_{24}& -S_{25}& S_{26}& -d_{12} & -d_{22}&  d_{32}\\
& & S_{33} & -S_{34}& -S_{35}& S_{36}& -d_{13} & -d_{23}&  d_{33}\\
& & & S_{44}& S_{45}& -S_{46}& d_{14} & d_{24}&  -d_{34}\\
& & & & S_{55}& -S_{56}& d_{15} & d_{25}&  -d_{35}\\
& &... & & & S_{66}& -d_{16} & -d_{26}&  d_{36}\\
& & & & & & \varepsilon_{11} & \varepsilon_{12}&  -\varepsilon_{13}\\
& & & & & & & \varepsilon_{22}&  -\varepsilon_{23}\\
& & & & & & & &  \varepsilon_{33}
\end{array}
\right), \qquad
{\bm q^s}^{\rm II}  = \left(
\begin{array}{r}
{\gamma_1^s}\\
{\gamma_2^s}\\
{\gamma_3^s}\\
-{\gamma_4^s}\\
-{\gamma_5^s}\\
{\gamma_6^s}\\
-{P_1^s}\\
-{P_2^s}\\
{P_3^s}
\end{array}
\right)
\]
In other words: the rotation $\bs \omega$ does not change the magnitude of the components of $\bm Q$ and $\bm q^{\rm s}$, but it changes the signs of some components:
\begin{alignat*}{3}
 Q_{\rm ij}^{\rm II} &= \abs{Q_{\rm ij}^{\rm I}} \qquad & {\rm i,j} &= 1..9 \\[1ex]
 Q_{\rm ij}^{\rm II} &= - Q_{\rm ij}^{\rm I} & {\rm i} &= 1,2,6,9; \pl &{\rm j} &= 4,5,7,8\\
 && {\rm i} &= 3; \pl &{\rm j} &= 4,5,7,8\\[1ex]
 {q_{\rm i}^{\rm s}}^{\rm II} &= \abs{ {q_{\rm i}^{\rm s}}^{\rm I} }, & {\rm i} &= 1..9 \\
 {q_{\rm i}^{\rm s}}^{\rm II} &= -{q_{\rm i}^{\rm s}}^{\rm I} , & {\rm i} &= 4,5,7,8  
\end{alignat*}
We note furthermore, that the submatrix $Q_{\rm \tilde i \tilde j}$ with $ \rm \tilde i, \tilde j = 1,2,6,9$ is homogeneous, i.e. according to the classification in the appendix, the domain system is partially homogenous. As it has been shown in \ref{detail_A}, internal fields and effective properties are considerably simplified in this case.
It follows from the condition of compatibility of the $\bm q^{\rm s}$ that
\begin{equation}
 {\bm p^{A}}^{\rm (k)} = \bm 0
\end{equation}
Then, \eqref{effprop_arb_e}-\eqref{effprop_arb_f} yield
\begin{equation} \label{q_sstern_twin}
q_{\rm i}^{s\, \ast} = \left\langle q_{\rm i}^{\rm s}\right\rangle, \quad {\rm i=1..9}\qquad  
\mbox{or} \qquad 
\bm q^{\rm s \ast} = \averg{\bm q^{\rm s }}
\end{equation}
It is convenient the to separate the interaction matrix $\bm A^{\rm (p)} = \bm I +\bm A'^{\rm (p)} $ 
with $\bf I$ beeing the 9$\times$9 identity matrix and
\begin{alignat}{3}
{A'}_{\rm \tilde i \tilde{\tilde{k}}}^{\rm (p)} &= \tilde{R}_{\rm \tilde i \tilde j} 
\left(\left\langle Q_{\rm \tilde{j} \tilde{\tilde{k}}}^{\rm (p)}\right\rangle 
- Q_{\rm \tilde{j} \tilde{\tilde{k}}}^{\rm (p)}\right) 
\qquad & \rm \tilde i &= 1,2,6,9; \quad &\rm \tilde{\tilde k } &= 3,4,5,7,8
\\[1ex] 
A_{\rm \tilde i \tilde k}^{\rm (p)} &= 0 & \rm \tilde i, \tilde k &= 1,2,6,9;
 \\[1ex]
{A'}_{\rm \tilde{\tilde{i}} k}^{\rm (p)} &= 0 \qquad & \rm \tilde{\tilde i} &=3,4,5,7,8; \quad & \rm k &= 1..9\\[2ex]
\end{alignat}
We may write for the internal fields
\[
\bm p^{\rm (p)} = \bm {\bar p} + \bm A' \bm {\bar p}
\]
In this representation the internal fields are expressed by a homogeneous contribution 
(corresponds to the generalized {\sc Reuss} approximation) and an additional contribution 
${A'}_{\rm ik}^{\rm Q\,(p)} \, \bar{p}_{\rm k}$
caused by the material inhomogeneity (anisotropy).

Additionally, the following components $Q_{\rm \tilde i \tilde{\tilde k}}$ are homogeneous:
\[
Q_{13},\pl Q_{23},\pl Q_{63},\pl Q_{93}
\]
and this yields $A_{\rm \tilde i \tilde{\tilde k}} = A_{13}= A_{23}= A_{63}= A_{93} = 0$, and
\[
\setlength{\arraycolsep}{7pt}
\renewcommand{\arraystretch}{1.2}
A_{ik}^{\rm (p)} = \left( 
\begin{array}{ccccccccc}
1 & 0 & 0 & A_{14}  & A_{15} & 0 & A_{17} & A_{18} & 0 \\
0 & 1 & 0 & A_{24}  & A_{25} & 0 & A_{27} & A_{28} & 0 \\
0 & 0 & 1 & 0            & 0           & 0 & 0           & 0      & 0 \\
0 & 0 & 0 & 1            & 0           & 0 & 0           & 0      & 0 \\
0 & 0 & 0 & 0            & 1           & 0 & 0           & 0      & 0 \\
0 & 0 & 0 & A_{64}  & A_{65} & 1 & A_{67} & A_{68} & 0 \\
0 & 0 & 0 & 0            & 0           & 0 & 1           & 0      & 0 \\
0 & 0 & 0 & 0            & 0           & 0 & 0           & 1      & 0 \\
0 & 0 & 0 & A_{94}  & A_{95} & 0 & A_{97} & A_{98} & 1 
\end{array}\right)
\]
It should be noted, that the interaction matrix $\bs A$ for a ferroelectric domain structure is an expression of the anisotropy of the linear material properties and of the crystallographic symmetry relationship between the two domains.

We can express the effective properties of a domain structure by the formula
\begin{alignat}{2}
\bm Q^{\ast} &= \averg{\bm Q} + \bm \Lambda \qquad \\
\intertext{with}
\Lambda_{\rm \tilde i k} &= 0 & \rm \tilde i &= 1,2,6,9 \\
& &\rm k &= 1..9 \nonumber \\[1ex]
\Lambda_{\rm \tilde {\tilde i} \tilde k} &= 0 & \rm \tilde {\tilde i} &= 3,4,5,7,8; \quad \\
& &\rm \tilde k& = 1,2,6,9 \nonumber \\[1ex]
\Lambda_{\rm \tilde{\tilde{i}} \tilde{\tilde{k}} } &=
\averg{ Q_{\rm \tilde{\tilde{i}}\tilde{j}}^{\rm (p)}}
   \tilde{R}_{\rm \tilde j \tilde m} 
\left\langle Q_{\rm \tilde{m} \tilde{\tilde{k}}}^{\rm (p)} \right\rangle 
- \averg{   Q_{\rm \tilde{\tilde{i}} \tilde{j}}^{\rm (p)} \tilde{R}_{\rm \tilde j \tilde m}
   Q_{\rm \tilde{m} \tilde{\tilde{k}}}^{\rm (p)}} \qquad & \rm \tilde{\tilde i}, \tilde{\tilde k} &= 3,4,5,7,8
\end{alignat}

As the $Q_{\rm i3}=Q_{\rm 3i} $, with i=1,2,6,9, are homogeneous, also the components $\Lambda_{\rm 3 i} = \Lambda_{i3} = 0$. The symmetric matrix $\Lambda_{\rm ik}$ has the following form:
\[
\setlength{\arraycolsep}{7pt}
\renewcommand{\arraystretch}{1.25}
\Lambda_{\rm ik}=
\left(\begin{array}{ccccccccc} 
0 & 0 & 0 & 0 & 0 & 0 & 0 & 0 & 0 \\
0 & 0 & 0 & 0 & 0 & 0 & 0 & 0 & 0 \\
0 & 0 & 0 & 0 & 0 & 0 & 0 & 0 & 0 \\
0 & 0 & 0 & \Lambda_{44} & \Lambda_{45} & 0 & \Lambda_{47} & \Lambda_{48} & 0  \\
0 & 0 & 0 & \Lambda_{45} & \Lambda_{55} & 0 & \Lambda_{57} & \Lambda_{58} & 0 \\
0 & 0 & 0 & 0 & 0 & 0 & 0 & 0 & 0 \\
0 & 0 & 0 & \Lambda_{47} & \Lambda_{57} & 0 & \Lambda_{77} & \Lambda_{78} & 0 \\
0 & 0 & 0 & \Lambda_{48} & \Lambda_{58} & 0 & \Lambda_{78} & \Lambda_{88} & 0 \\
0 & 0 & 0 & 0 & 0 & 0 & 0 & 0 & 0 
\end{array} \right)
\]Therefore, in a ferroelectric domain structure with arbitrary symmetry, the only components of the effective linear properties, which differ from the simple average $\averg{\bm Q} $ (lower bound) are $Q_{44}=S_{44}$, $Q_{45}=S_{45}$, $Q_{47}=d_{14}$, $Q_{48}=d_{24}$, $Q_{55}=S_{55}$, $Q_{57}=d_{15}$, $Q_{58}=d_{25}$, $Q_{77}=\varepsilon_{11}$, $Q_{78}=\varepsilon_{12}$ and $Q_{88}=\varepsilon_{22}$.

The constants $S_{11}$,
$S_{12}$,
$S_{13}$,
$S_{16}$,
$S_{22}$,
$S_{23}$,
$S_{33}$,
$S_{26}$,
$S_{66}$,
$d_{31}$,
$d_{32}$,  
$d_{36}$ 
and $\varepsilon_{33}$,
of the single crystal, are homogenous and the effective constants are equal to the single crystal values. 

The effective constants 
%AB
$S_{14}^\ast$,
$S_{15}^\ast$,
$S_{24}^\ast$,
$S_{25}^\ast$,
$S_{34}^\ast$,
$S_{35}^\ast$,
$S_{36}^\ast$,
$S_{46}^\ast$,
$S_{56}^\ast$,
$d_{11}^\ast$,
$d_{21}^\ast$,
$d_{12}^\ast$,
$d_{22}^\ast$,
$d_{13}^\ast$,
$d_{23}^\ast$,
$d_{33}^\ast$,
$d_{34}^\ast$,
$d_{35}^\ast$,
$d_{16}^\ast$,
$d_{26}^\ast$,
$\varepsilon_{13}^\ast$ and 
$\varepsilon_{23}^\ast$ depend linearly upon the volume fraction and are zero at $\xi=0.5$.

It should be emphasized, that the simplifications for ferroelectric domain structures, which have been derived in this section, are not possible for alternative sets of effective constants, like $\bm R^\ast$ or $\bm L^\ast$. The internal field components, which have to be continuous across the interface are only  for applied stresses and electric fields equal to the macroscopic fields. Alternative set of constants have to be calculated using formulas similar to those derived in sect. \ref{arbitrary_laminate}.

In the following we want to study important cases for particular crystal symmetries.

\subsection{Tetragonal $90^\circ$ domain structures}\label{tetrag-90}

Barium titanate ($\rm BaTiO_3$), lead titanate ($\rm PbTiO_3$) and titantium rich lead zirconate titanate ($\rm Pb Zr_xTi_{1-x}O_3$, PZT) are important tetragonal piezo- and ferroelectric material, which often show laminar 90$^\circ$ domain structures, as sketched in fig. \ref{domaenen_1}.

The properties of the single domain single crystal are usually described in a crystal coordinate system with the tetragonal axis (pseudo-cubic [001] = polarization direction) parallel to macroscopic $x_3$ \citep{nye}. We will denote the single crystal properties described in the crystal coordinate system by the suffix $^\circ$: ${\bm Q}^\circ$, ${\bm q^{\rm s}}^\circ$ 
A tetragonal ferroelectric phase is characterized by spontaneous polarization and strain 
\[
{\bs P^{\rm s}}^{\circ} = \left(\begin{array}{r} 0  \\ 0  \\ P_3^{\rm s}    
\end{array}
\right), 
\qquad 
\setlength{\arraycolsep}{5pt}
{\bs \gamma^{\rm s}}^\circ = \left( \begin{array}{rrr}
\gamma_{11} & 0   & 0   \\
& \gamma_{11}   &0   \\
...& &  \gamma_{33}  
\end{array}\right)
\]
There are six possible tetragonal variants, characterized by the polarization directions: [001], [100], [010] and their negative directions, and three spontaneous strain states associated with the tetragonal phase transformation.
We may distinguish two principle domain types: 90$^\circ$ domains, where the angle between the polarization directions is approximately 90$^\circ$, the strain states are different and the domain walls are the pseudo-cubic $\{110 \}$ planes. The second type of tetragonal domains are 180$^\circ$ domains, which have identical strain in both domains, but the polarizations point in opposite directions. All planes with a normal vector perpendicular to the polarization are possible domain walls. This example is considered in sect. \ref{tetrag-180}.

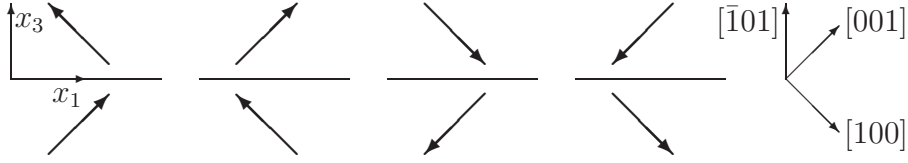
\begin{figure}[htbp]
\begin{center}
\begin{picture}(12,2)
%\put(0,0){\framebox(12,2){}}
\put(0,0){
\put(0,1){\vector(1,0){1}}\put(0,1.5){\makebox(0.5,0.5){$x_3$}}
\put(0,1){\vector(0,1){1}}\put(0.5,0.5){\makebox(0.5,0.5){$x_1$}}
}

\put(0,0){
\put(0,1){\line(1,0){2}}
\thicklines
\put(0.5,0){\vector(1,1){0.8}}
\put(1.3,1.2){\vector(-1,1){0.8}}}

\put(2.5,0){\put(0,1){\line(1,0){2}}
\thicklines
\put(1.3,0){\vector(-1,1){0.8}}
\put(0.5,1.2){\vector(1,1){0.8}}}

\put(5,0){\put(0,1){\line(1,0){2}}
\thicklines
\put(1.3,0.8){\vector(-1,-1){0.8}}
\put(0.5,2){\vector(1,-1){0.8}}}

\put(7.5,0){\put(0,1){\line(1,0){2}}
\thicklines
\put(0.5,0.8){\vector(1,-1){0.8}}
\put(1.3,2){\vector(-1,-1){0.8}}}

\put(10.3,1){\vector(0,1){1}} \put(9.3,1.5){\makebox(1,0.5){$[\bar 101]$}}
\put(10.3,1){\vector(1,1){0.71}} \put(11,1.5){\makebox(1,0.5){$[001]$}}
\put(10.3,1){\vector(1,-1){0.71}} \put(11,0){\makebox(1,0.5){$[100]$}}

\end{picture}
\caption{The four possible 90$^\circ$ domain structures with a $(\bar 1 0 1) || \bs n=(0,0,1)^t $ domain wall.}
\label{possible-90}
\end{center}
\end{figure}

We assume, that the single crystal in the pseudo-cubic parelectric state is rotated by 45$^\circ$ about the $x_2$ axis,  so that the pseudo-cubic $(\bar 1 01)$ becomes parallel to the plane with $\bs n = (0,0,1)^t$
There are four possible arrangements of tetragonal domains with this plane as a domain wall (figure \ref{possible-90}).

For the purpose of this work, it is sufficient to consider only the first example. The second one is easily obtained by interchanging the volume fractions of domain I and II, and the remaining two arrangements are the same as the first two, but with inverse polarization directions (rotated by 180$^\circ$).

In order to end up with a compatible spontaneous strain, the domain I is rotated the angle $\theta$ \citep{shu.01}
\[
\theta = \frac{\pi}{2} + \phi, 
\qquad \mbox{with} \quad
\phi = \arccos \left(  \frac{2 \gamma_{11}^{\rm s^\circ}  \gamma_{33}^{\rm s^\circ}}{  {\gamma_{11}^{\rm s^\circ}}^2 +  {\gamma_{11}^{\rm s^\circ} }^2 }  \right).
\]
As the spontaneous strain is small the rotation $\phi$ is small and has negligible effects on the effective properties.
The orientation of domain II may be obtained from domain I by applying the orientation relationship \eqref{orient-relation} and \eqref{rotation_comptbl}.

The, we find 
$\gamma_{44}^{\rm s},\, \gamma_{66}^{\rm s} =0 $ and 
$P_{2}^{\rm s} = 0$ in both domains.
For the rotated linear properties holds
$\varepsilon_{12}$, $\varepsilon_{23}$,
$d_{14}$, $d_{16}$, $d_{21}$, $d_{22}$, $d_{23}$, $d_{25}$, $d_{34}$, $d_{36}$,  
$S_{14}$, $S_{16}$, $S_{24}$, $S_{26}$, $S_{34}$, $S_{36}$, $S_{45}$, $S_{56}=0$ in both domains.
%$\varepsilon_{12} = \varepsilon_{23} = 0$, 
%$d_{14}=d_{16}=d_{21}=d_{22}=d_{23}=d_{25}=d_{34}=d_{36}=0$,
%$S_{14}=S_{16}=S_{24}=S_{26}=S_{34}=S_{36}=S_{45}=S_{56}=0$ in both domains.

This results in further components of the interaction matrix $\bs A$ to be zero:
\begin{eqnarray*}
A_{14}^{\rm (p)} = A_{24}^{\rm (p)} = A_{94}^{\rm (p)} = 0\\
A_{65}^{\rm (p)} = A_{67}^{\rm (p)} = 0\\
A_{18}^{\rm (p)} = A_{28}^{\rm (p)} = A_{98}^{\rm (p)} = 0,
\end{eqnarray*}

Then follows that for tetragonal 90$^\circ$ domains the interaction matrix $\bf A^{\rm (p)}$ has the form:
\[
\setlength{\arraycolsep}{7pt}
\renewcommand{\arraystretch}{1.2}
A_{ik}^{\rm (p)} = \left( 
\begin{array}{ccccccccc}
1 & 0 & 0 & 0      & A_{15} & 0 & A_{17} & 0      & 0 \\
0 & 1 & 0 & 0      & A_{25} & 0 & A_{27} & 0      & 0 \\
0 & 0 & 1 & 0      & 0      & 0 & 0      & 0      & 0 \\
0 & 0 & 0 & 1      & 0      & 0 & 0      & 0      & 0 \\
0 & 0 & 0 & 0      & 1      & 0 & 0      & 0      & 0 \\
0 & 0 & 0 & A_{64} & 0      & 1 & 0      & A_{68} & 0 \\
0 & 0 & 0 & 0      & 0      & 0 & 1      & 0      & 0 \\
0 & 0 & 0 & 0      & 0      & 0 & 0      & 1      & 0 \\
0 & 0 & 0 & 0      & A_{95} & 0 & A_{97} & 0      & 1 
\end{array}\right).
\]

\subsection{Tetragonal $180^\circ$ domain structures}\label{tetrag-180}

\begin{figure}[htbp]
\begin{center}
\begin{picture}(7,5.5)
\put(0,0){\includegraphics[width=7cm]{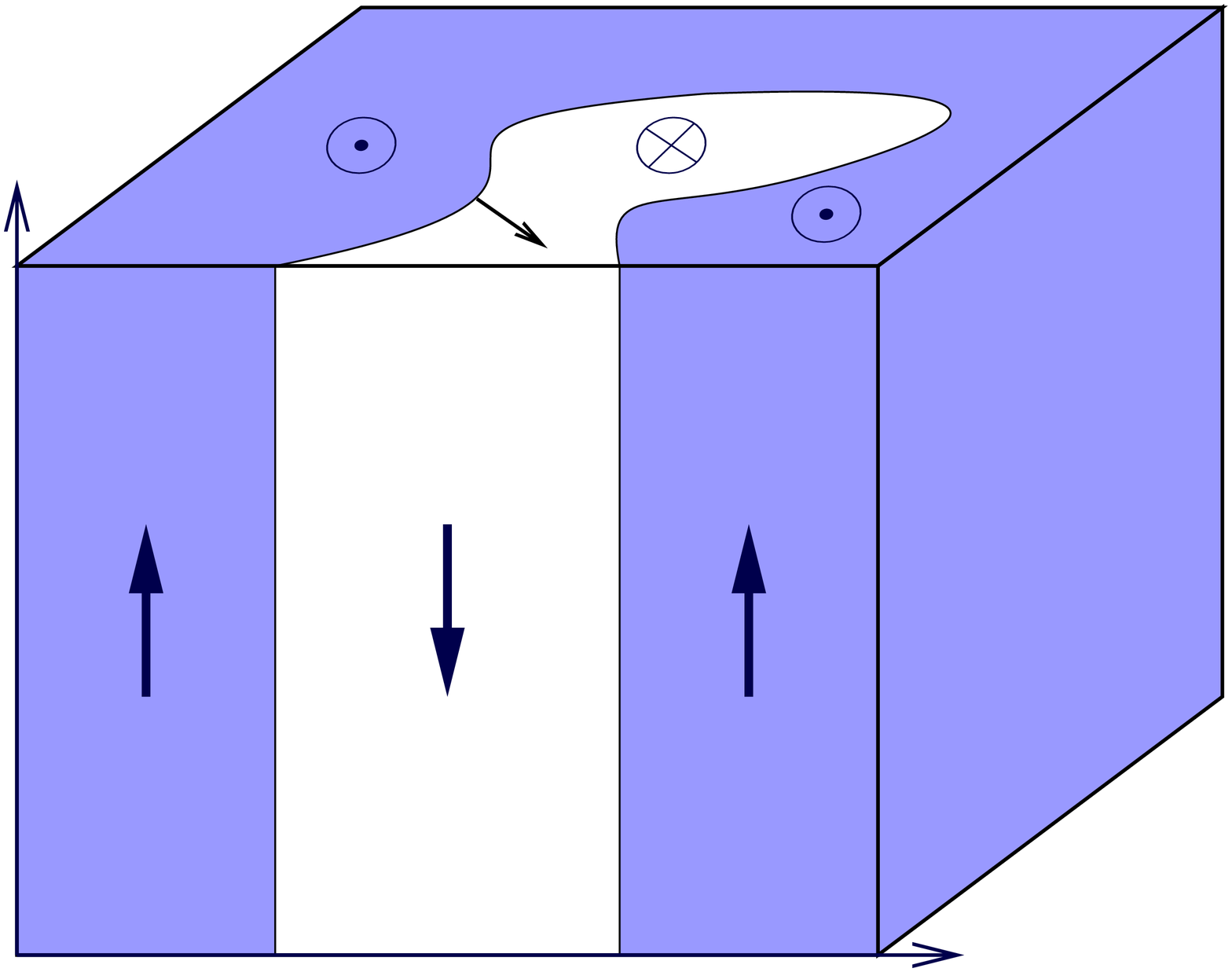}}
\put(5.3,0.3){[100]}
\put(0,4.5){[001]}
\put(3,4.2){$\bs n$}
\put(0,5.2){a)}
\end{picture}
\begin{picture}(10,7)
%\put(0,0){\framebox(10,7){}}
\put(0,0){\includegraphics[width=10cm]{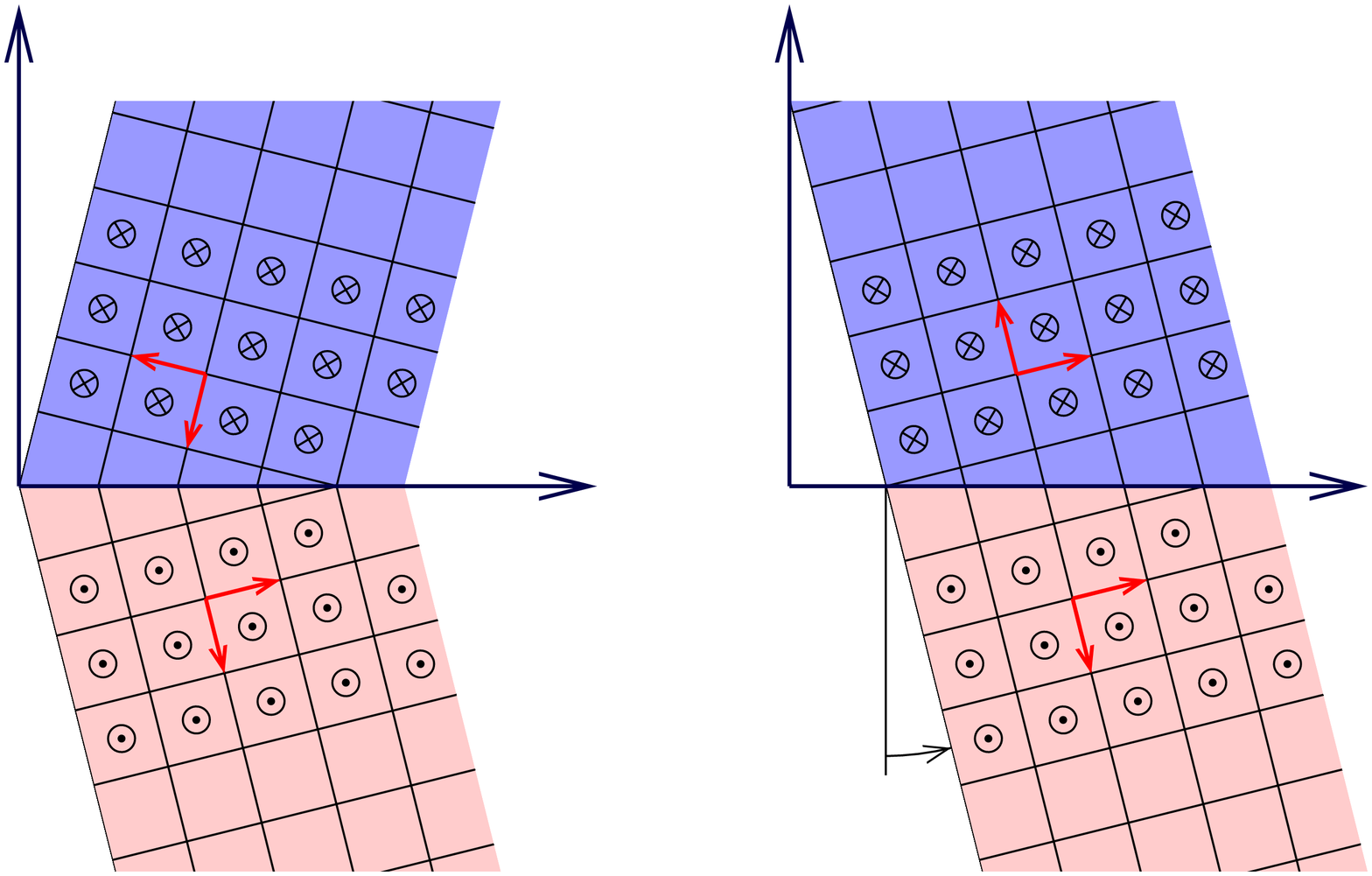}}
\put(0,6.5){b)}
\put(5,6.5){c)}
\put(0.8,6){$x_3$}
\put(4,2.7){$x_2$}
\put(5.8,6){$x_3$}
\put(9.5,2.7){$x_2$}
\put(6.2,1){$\psi$}
\end{picture}
\caption{180$^\circ$ domains. a) Interfaces between compatible, anti-parallel domains are not necessarily parallel to simple crystallographic planes. 
b) Compatible bi-crystal with tilted crystal lattices but anti-parallel polarizations. c) 180$^\circ$ domain wall with anti-parallel crystal lattices vectors, but no tilt.}
\label{DW-180}
\end{center}
\end{figure}

A particular simple case of an 180$^\circ$ domain structure is obtained, assuming that the pseudo-cubic (100) plane is the domain wall. However, we want to consider the general case, where all planes with $\bs n \perp \bs P$ are allowed (see fig. \ref{DW-180} a). Then, an additional rotation $\psi$ about the [001]-axis of the crystal should be taken into account and the transformation matrix, $\bs \omega'$, which rotates the reference orientation into an arbitrarily oriented domain I, is: 
\[
\bs x^{\rm I} = \bs \omega' \, \bs x^\circ,
\qquad
\setlength{\arraycolsep}{7pt}
\bs \omega' = \left(\begin{array}{ccc} 0&0&1\\
\sin \psi &\cos \psi&0\\ -\cos \psi& \sin \psi &0\end{array} \right).
\]
Then, we find the crystal properties of domain I with respect to the macroscopic coordinate system in terms of the single crystal properties in crystal reference coordinates, $\bs S^\circ$,  $\bs d^\circ$,  $\bs \varepsilon^\circ$, $\bs \gamma^{\rm s\,\circ}$, and $\bs P^{\rm s\,\circ}$:  
\[
\setlength{\arraycolsep}{5pt}
\bm Q^{\rm I}=
\left(
\begin{array}{ccr rcc ccc}
S_{33}^\circ  &  S_{13}^\circ  & S_{13}^\circ  & 0\;\; & 0 & 0 & d_{33} ^\circ & 0 &  0 \\
S_{13}^\circ  &  S'_{22} & S'_{23} & S'_{24} & 0 & 0 & d_{31}^\circ  & 0 & 0 \\
S_{13}^\circ  &  S_{23}^\circ  & S'_{22}  & -S'_{24} & 0 & 0 & d_{31}^\circ  & 0 & 0 \\
0  &  S'_{24}  & -S'_{24}  &  S'_{44} & 0 & 0 & 0 & 0 & 0 \\
0  &  0  & 0\;\;  & 0\;\; & S_{44}^\circ & 0 & 0 & 0 &  d_{15}^\circ \\
0  &  0  & 0\;\;  & 0\;\; & 0 & S_{44}^\circ & 0  & d_{15}^\circ &  0 \\
d_{33}^\circ  &  d_{31}^\circ  & d_{31}^\circ  & 0\;\; & 0 & 0 & \varepsilon_{33}^\circ  & 0 & 0\\
0 & 0 & 0\;\; & 0\;\; & 0 & d_{15}^\circ & 0 & \varepsilon_{11}^\circ &  0 \\
0 & 0 & 0\;\; & 0\;\; & d_{15}^\circ & 0 & 0 & 0 &  \varepsilon_{11}^\circ 
\end{array}
\right), \quad
\bm q^{\rm s\, I} = \left(
\begin{array}{c}
{\gamma_3^{\rm s\, \circ}}\\
{\gamma_1^{\rm s\, \circ}}\\
{\gamma_1^{\rm s\, \circ}}\\
0\\
0\\
0\\
{P_3^{\rm s\, \circ}}\\
0\\
0
\end{array}
\right)
\]
\begin{align*}
S' &=  [2 (S_{11}^\circ-S_{12}^\circ) - S_{66}^\circ)]  \\
S'_{22} &= S_{12}^\circ - (\cos^4 \psi -\cos^2 \psi) \, S' \\
S'_{23} &= S_{11}^\circ + (\cos^4 \psi -\cos^2 \psi) \, S' \\
S'_{24} &= - \frac{1}{4} \sin (4 \psi) \,  S' \\
S'_{44} & =  S_{66}^\circ-4 (\cos^4 \psi -\cos^2 \psi) \, S' 
\end{align*}

In the case of a 180$^\circ$ domain it is interesting to note, that applying the orientation relationship \eqref{orient-relation} and \eqref{rotation_comptbl}, will yield two anti-parallel oriented domains. However, the resulting structure (fig. \ref{DW-180} b) is generally not a 180$^\circ$ domain structure in the common sense, because the crystal lattices of the two domains are tilted with respect to each other.
Although this does not violate the condition of compatibility of strain and polarization, we should denote this example rather as a {\em bi-crystal} then a domain structure.
A domain structure with anti-parallel polarizations, which forms as a result of a symmetry breaking phase transformation (fig. \ref{DW-180} c) can be described by following transformation matrices:
\[
\bs x^{\rm II} = \bs \omega'' \, \bs x^\circ,
\qquad
\setlength{\arraycolsep}{7pt}
\bs \omega'' = \left(\begin{array}{ccc} 0&0&-1\\
-\sin \psi &\cos \psi&0\\ \cos \psi& \sin \psi &0\end{array} \right).
\]

It is obvious, that the elastic and dielectric properties of the two domains are identical, but all piezoelectric coefficients will have opposite signs. $\bs {\tilde R}$ is again homogeneous.

We find, that the following components of the interaction matrix $\bm A$ become zero:
$A_{14}, A_{24}, A_{64} , A_{94}, A_{15}, A_{25}, A_{65}, A_{67}, A_{97}, A_{18}, A_{28}, A_{98} = 0$,
and the remaining components are 
$ A_{95}, A_{17}, A_{27}, A_{68} \ne 0$.
\[
\setlength{\arraycolsep}{7pt}
\renewcommand{\arraystretch}{1.2}
A_{ik}^{\rm Q\,(p)} = \left( 
\begin{array}{ccccccccc}
1 & 0 & 0 & 0 & 0 & 0 & A_{17} & 0      & 0 \\
0 & 1 & 0 & 0 & 0 & 0 & A_{27} & 0      & 0 \\
0 & 0 & 1 & 0 & 0      & 0 & 0      & 0      & 0 \\
0 & 0 & 0 & 1 & 0      & 0 & 0      & 0      & 0 \\
0 & 0 & 0 & 0 & 1      & 0 & 0      & 0      & 0 \\
0 & 0 & 0 & 0 & 0      & 1 & 0      & A_{68} & 0 \\
0 & 0 & 0 & 0 & 0      & 0 & 1      & 0      & 0 \\
0 & 0 & 0 & 0 & 0      & 0 & 0      & 1      & 0 \\
0 & 0 & 0 & 0 & A_{95} & 0 & 0  & 0      & 1 
\end{array}\right)
\]

The following components of $\bs \Lambda$ are zero:  $\Lambda_{44}$, $\Lambda_{45}$,  $\Lambda_{47}$, $\Lambda_{48}$, $\Lambda_{57}$, $\Lambda_{58}$ and $\Lambda_{78}$.   
The matrix $\bs \Lambda$ takes the form:
\[
\setlength{\arraycolsep}{5pt}
\Lambda_{\rm ik}=
\left(\begin{array}{ccccccccc} 
0 & 0 & 0 & 0 & 0 & 0 & 0 & 0 & 0 \\
0 & 0 & 0 & 0 & 0 & 0 & 0 & 0 & 0 \\
0 & 0 & 0 & 0 & 0 & 0 & 0 & 0 & 0 \\
0 & 0 & 0 & 0 & 0 & 0 & 0 & 0& 0  \\
0 & 0 & 0 & 0 & \Lambda_{55} & 0 & 0 & 0& 0 \\
0 & 0 & 0 & 0 & 0 & 0 & 0 & 0 & 0 \\
0 & 0 & 0 & 0 & 0 & 0 & \Lambda_{77} & 0 & 0 \\
0 & 0 & 0 & 0 &  0 & 0 & 0 & \Lambda_{88} & 0 \\
0 & 0 & 0 & 0 & 0 & 0 & 0 & 0 & 0 
\end{array} \right)
\]

As a result, the effective properties of a 180$^\circ$ domain structure is very close to the generalized Reuss approximation.The three contributions $\Lambda_{55}$, $\Lambda_{77}$ and $\Lambda_{88}$ can easily be explained. 
The piezoelectric shear effect will cause opposed dielectric displacements, if a macroscopic shear stress $\bar \sigma_5$ is applied. This results in a reduced compliance of the composite.
Anti directed in-plane shear of the two domains in the case of applied $\bar E_2$ due to the inverse piezoelectric effect results in in-plane shear stresses which restrict the free deformation of the stack. This results in a decrease of effective $\varepsilon_{22}^{\sigma \,\ast}$.
If the volume fractions are equal, the effective dielectric constant in 2-direction, $\varepsilon_{22}^{\sigma\,\ast}$, equals the single crystal dielectric constant under fixed strain $\varepsilon_{22}^{\gamma\, \ast}$.
A similar effect influences the dielectric constant in polarization direction, $\varepsilon_{11}^\ast$, where opposed elongation and compression in the two domains are caused by an applied $\bar E_1$.
This effect is, of course, also present for other domain orientations, but most easily explained in the 180$^\circ$ case.

For an 180$^\circ$ bi-crystal (fig. \ref{DW-180} b), $S_{24}$ and $S_{34}$ will have opposite signs. Then, the components $A_{14}$ and $A_{24}$ will not be zero. In this case, an applied shear stress $\bar \sigma_4$ will cause inhomogeneous internal stresses $\sigma_1$ and $ \sigma_2$.
Moreover, $\Lambda_{44}$ and $\Lambda_{47}$ will be non-zero because due to the piezoelectric coupling, the inhomogeneous stresses $\sigma_1$ and $ \sigma_2$ will result in dielectric displacements $D_1$.
This is particular interesting, as it means, that there may appear an additional piezoelectric coefficient $d_{14}^\ast$ for the case of the bi-crystal, although the coefficients of the uncoupled constituents are zero.

\newpage

\subsection{Rhombohedral domains}\label{rh-domains}

Within the framework described above, it is easy to analyze other crystal symmetries. In particular, the treatment of rhombohedral domain structures is very similar to the tetragonal case.
Rhombohedral phases of space group {\em R3m} and {\em R3c} appear in many solid solutions of lead titanate, like $\rm Pb Zr_xTi_{1-x}O_3$ with $x>0.52$, or PMN-$x$PT ($\rm Pb(Mg_{1/3}Nb_{2/3})_x Ti_{1-x} O_3$) with $x>0.7$.
The rhombohedral phase is characterized by a spontaneous polarization along the pseudo-cubic [111] direction and a spontaneous strain which results in the rhombohedral distortion of the unit cell:
\[
\bs P^{\rm s} = \frac{1}{\sqrt{3}}\left(
\begin{array}{c}
P^s \\ P^s  \\ P^s   
\end{array}
\right), 
\qquad
\bs \gamma^{\rm s} = \left(
\begin{array}{ccc}
\gamma_{11}^{\rm s}   & \gamma_{13}^{\rm s}   & \gamma_{13}^{\rm s}   \\
\gamma_{13}^{\rm s}   & \gamma_{11}^{\rm s}   & \gamma_{13}^{\rm s}   \\
\gamma_{13}^{\rm s}   & \gamma_{13}^{\rm s}   & \gamma_{11}^{\rm s}  
\end{array}
\right)
\]

\begin{figure}[htbp]
\begin{center}
\begin{picture}(12,6.5)
%\put(0,0){\framebox(12,6.5){}}
\put(1,1){\vector(1,0){3.5}}
\put(1,1){\vector(0,1){4.5}}
\put(4.2,0.5){\makebox(0,0)[l]{[110]}}
\put(0.5,5.7){\makebox(0,0)[l]{[001]}}
\put(3,4.2){\makebox(0,0)[l]{[$\bar 1 \bar 11$]}}
\put(3,1.8){\makebox(0,0)[l]{[111]}}

\put(1,1){\framebox(3,2){}}
\put(1,1){\vector(3,2){3}}
\put(1,3){\framebox(3,2){}}
\put(4,3){\vector(-3,2){3}}
\put(2.,5.5){\makebox(0,0)[l]{70.53$^\circ$}}
\put(1.5,2.5){\makebox(0,0){I}}
\put(1.5,3.5){\makebox(0,0){II}}

\put(5,2){\vector(1,0){6.5}}
\put(5,2){\vector(0,1){2.5}}
\put(11,1.5){\makebox(0,0)[l]{[110]}}
\put(4.5,4.9){\makebox(0,0)[l]{[001]}}
\put(7,3.2){\makebox(0,0)[l]{[111]}}
\put(10,2.8){\makebox(0,0)[l]{[11$\bar 1$]}}

\put(8,4){\vector(3,-2){3}}
\put(8,2){\framebox(3,2){}}
\put(5,2){\vector(3,2){3}}
\put(5,2){\framebox(3,2){}}
\put(7.5,4.5){\makebox(0,0)[l]{109.47$^\circ$}}
\put(5.5,3.5){\makebox(0,0){I}}
\put(8.5,2.5){\makebox(0,0){II}}

\end{picture}
\caption{Simple non-180$^\circ$ domain patterns in rhombohedral phases}
\label{rh-71-109}
\end{center}
\end{figure}
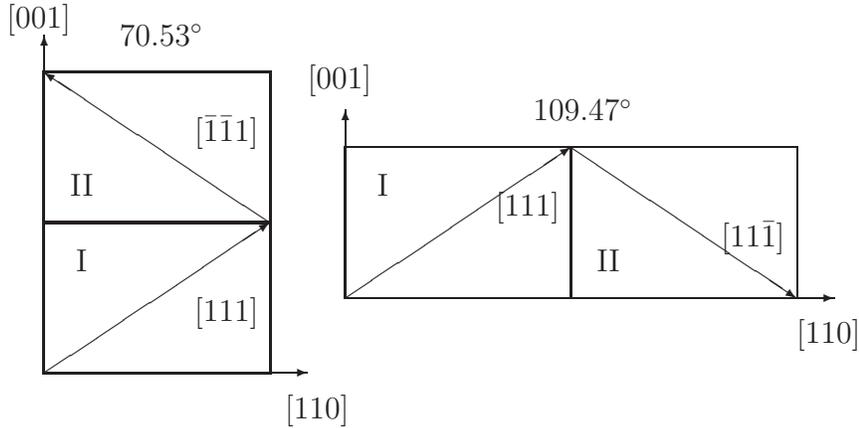

There exist two types of non-180$^\circ$ domain patterns, the first one, has the (001) plane as the domain interface and the angle between the polarization vectors  $\bs P^{\rm s \, I}$ and $- \bs P^{\rm s \, II}$ is $\arccos \frac{1}{3} \approx 70.53^\circ$. In a second type of domain structures the (110) plane is the domain interface and the angle between $\bs P^{\rm s \, I}$ and $- \bs P^{\rm s \, II}$ is
$\arccos \left(  -\frac{1}{3}  \right) \approx 109.47^\circ$.
This principal patterns may are sketched in $(1 \bar 1 0)$ plane in fig. \ref{rh-71-109}.

Due to the rhombohedral distortion, a small additional rotation by angle $\phi$ about the pseudo-cubic axis $[\bar 1 1 0] $ is necessary to ensure strain compatibility. For the case of a (001) domain interface $\phi$  is 
\[
\phi = \arccos \left(  \frac{\gamma_{11}^{\rm s} + \gamma_{13}^{\rm s}}{\sqrt{{\gamma_{11}^{\rm s}}^2 + 2 \gamma_{11}^{\rm s}\gamma_{13}^{\rm s} + 3{\gamma_{13}^{\rm s}}^2}}  \right),
\]
and for the (110) interface $\phi$ is
\[
\phi = - \arccos \left(  \frac{\gamma_{11}^{\rm s} }{\sqrt{{\gamma_{11}^{\rm s}}^2 + 2{\gamma_{13}^{\rm s}}^2}}  \right).
\]
Again, as the spontaneous strain is small the rotation $\phi$ is small and has negligible effects on the effective properties.

\begin{figure}[htbp]
\begin{center}
\begin{picture}(10,8.5)
%\put(0,0){\framebox(10,8.5){}}
\put(0,0){\includegraphics[width=10cm]{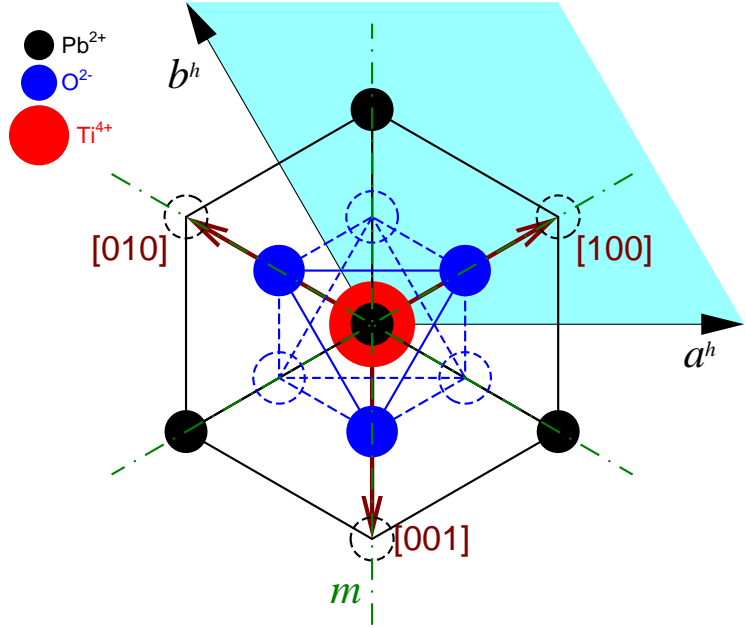}}
\end{picture}
\caption{Perovskite unit cell in hexagonal (trigonal) setting, seen along the pseudo-cubic [111] direction (hexagonal c axis). The corresponding hexagonal unit cell is indicated by the lattice vectors $\bs a^h$ and $\bs b^h$. The orientation of the pseudo-cubic cell is indicated by pseudo-cubic lattice vectors [100], [010] and [001].}
\label{rh-hex}
\end{center}
\end{figure}

The single crystal properties of a rhombohedral material are usually described in the so-called hexagonal setting, where the pseudo-cubic [111] axis is the crystal c-axis and parallel to the macroscopic $x_3$ direction and the crystal has trigonal symmetry (\citet{nye}, fig. \ref{rh-hex}). With respect to this orientation, the linear property matrix of a rhombohedral single crystal, $\bs Q^h$ and $\bs q^{\rm s\, h}$is:
\[
\bs Q^h = \left(
\begin{array}{rrc rrr rrc}
S_{11}^h  &  S_{12}^h  & S_{13}^h  & S_{14}^h & 0\;\, & 0\;\, & 0\;\, & -d_{22}^h  &  d_{31}^h \\
S_{12}^h  &  S_{11}^h & S_{13}^h & -S_{14}^h & 0\;\, & 0\;\, & 0\;\,  & d_{22}^h  & d_{31}^h \\
S_{13}^h  &  S_{13}^h  & S_{33}^h  & 0\;\, & 0\;\, & 0\;\, & 0\;\, & 0\;\, & d_{33}^h  \\
S_{14}^h  &  -S_{14}^h  & 0  &  S_{44}^h & 0\;\, & 0\;\, & 0\;\, & d_{15}^h & 0 \\
0\;\,  &  0\;\,  & 0  & 0\;\, & S_{44}^h & 2S_{14}^h & d_{15}^h & 0\;\, &  0 \\
0\;\,  &  0\;\,  & 0  & 0\;\, & 2S_{14}^h & S_{66}^h & -2d_{22}^h & 0\;\, &  0 \\

0\;\,  &  0\;\,  & 0  & 0\;\, & d_{15}^h & -2d_{22}^h  & \varepsilon_{11}^h & 0\;\, & 0\\
-d_{22}^h   & d_{22}^h & 0 & d_{15}^h & 0\;\, & 0\;\, & 0\;\, & \varepsilon_{11}^h &  0 \\
d_{31}^h & d_{31}^h & d_{33}^h & 0\;\, & 0\;\,  & 0\;\, & 0\;\, & 0\;\, &  \varepsilon_{33}^h 
\end{array}
\right), \quad
\bm q^{\rm s\, I} = \left(
\begin{array}{c}
{\gamma_1^{\rm s\, h}}\\
{\gamma_1^{\rm s\, h}}\\
{\gamma_3^{\rm s\, h}}\\
0\\
0\\
0\\
0\\
0\\
{P_3^{\rm s\, h}}
\end{array}
\right)
\]
\[
S_{66}^h = 2 (S_{11}^h-S_{12}^h)
\]

First, we want to consider a $70.53^\circ$ domain structure with (001) interface.  In order to apply the formalism of sect \ref{compatible-rank-1} we have to rotate the unit cell by $-90^\circ$ about the $x_3$ axis and by an angle $\arccos (1/\sqrt{3})  \approx 55^\circ $ about the 
$x_2$ axis, yielding the orientation of domain I with the pseudo-cubic [110] direction parallel to $x_1$.
\[
\bs x^{\rm I} = \bs \omega'  \bs x^h, \qquad \bs \omega' = 
\left(
\begin{array}{r cc}
 0 &  \frac{1}{\sqrt{3}} & \sqrt{\frac{2}{3}}  \\
 -1 & 0  & 0  \\
 0 &  -\sqrt{\frac{2}{3}} &   \frac{1}{\sqrt{3}} 
\end{array}
\right)
\]

Similarily, in order to describe the 109$^\circ$ domain structure with the (110) interface the rotation $\bs \omega ''$ will rotate the domain I so that $\bs n = (0,0,1)^t$ is the interface normal to the pseudo-cubic (110)  plane.
\begin{equation}
\bs x^{\rm I} = \bs \omega''  \bs x^h, \qquad \bs \omega''= 
\left(
\begin{array}{r cc}
0 &  -\sqrt{\frac{2}{3}} &   \frac{1}{\sqrt{3}} \\
0 &  \frac{1}{\sqrt{3}} & \sqrt{\frac{2}{3}}   \\
 -1 & 0  & 0 
\end{array}
\right) \label{rh-109-orientation}
\end{equation}

Alternatively, the same result is obtained by subsequent rotations of the domain I in the 71$^\circ$ setting by -90$^\circ$ and 180$^\circ$ about the $x_2$- and $x_3$-axis, respectively.
Then, orientation relationship \eqref{orient-relation} and \eqref{rotation_comptbl} can be applied in order to get the orientation and properties of domain II for both cases.
%The resulting lattice orientations are in agreement with the requirements of domain formations due phase transformations.
The effective properties of the rhombohedral domain pattern can be calculated using formulas of sect. \ref{compatible-rank-1}.

Note, that the chosen orientation of domain I is in principle a monoclinic or othorhombic setting of the unit cell, i.e. the reference orientation of monoclinic or othorhombic  crystals can be directly used to calculate the effective properties of monoclinic domain structures.

\newpage

%%%%%%%%%%%%%%%%%%%%%%%%%%%%%%%%%%

\newpage

\section{Charged domain walls}

Although, compatible domain structures are considered to be favored for sake of energy minimization, the existence of not fully compatible structures are quite common. For instance by applying sufficiently high electric fields to a single crystal, the domains can arrange themselves so that the spontaneous strains are compatible but the polarizations are not. This is possible, if the average polarization of all domains, which are favored by the direction of the applied field, is aligned with the field, but there is no possibility to arrange them so that the polarizations are compatible.
For instance, in a tetragonal crystal, an applied electric field in [111]-direction would favor the [100], [010], and [001]-domain, but there is no possibility to arrange this polarizations in a way, that gives continuous polarizations across all domain boundaries. However, the strains can be compatible. The concept of domain engineering makes use of this principle.
We will study this particular example in detail in sect. \ref{tetragonal_engineered_111}.  
In this section we want to emphasize the differences between a fully compatible rank-1 laminate (sect. \ref{compatible-rank-1}) and a rank-1 laminate with incompatible polarizations for arbitrary material symmetry.

\begin{figure}[hbtp]
\begin{center}
\begin{picture}(10,8)
\put(0,0){\includegraphics[width=10cm]{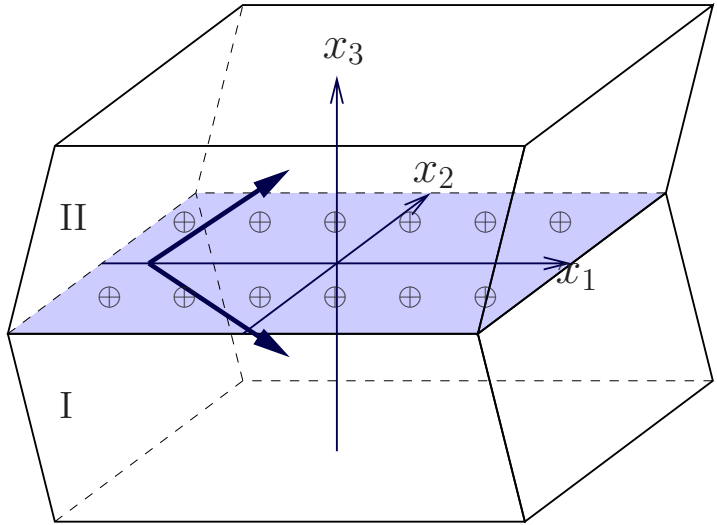}}
\put(7.6,3.8){\Large$ x_1$}
\put(5.7,5.15){\Large $x_2$}
\put(4.5,6.8){\Large$ x_3$}
\put(1,2){\large I }
\put(1,4.5){\large II}
\multiput(1.5,3.5)(1.0,0){6}{\multiput(0,0)(1.0,1.){2}{$\oplus$}}
\end{picture}
\caption{\label{chargedwall} Orientation relationship of compatible ferroelectric  domains}
\end{center}
\end{figure}

Consider a domain I, arbitrarily oriented with respect a plane with normal  $\bs n= (0,0,1)^t$, as in sect. \ref{compatible-rank-1} (fig. \ref{chargedwall}).
A domain II with compatible strains can be created by rotating the material on the other side of the interface by 180$^\circ$ about the $x_1$-axis.
Then, the rotation $\bs \omega$ 
\begin{equation}
\bs \omega = \left(
\begin{array}{rrr}
 1 & 0  & 0  \\
  0 & \; -1 & \pl 0  \\
  0 & 0  & -1  
\end{array}
\right)
\label{rotation_charged}
\end{equation}
yield a mechanically compatible, but electrically incompatible domain structure, if the strain component
$\gamma_{12}^{\rm s} = 0$. This means, in order to describe compatible spontaneous strains, we have to chose the direction of the coordinate system so that  the in-plane shear vanishes, which should be always possible.

We obtain the following properties of domain II:
\[
\bm Q^{\rm II}=
\left(
\begin{array}{rrr rrr rrr}
S_{11} &  S_{12} & S_{13} &  S_{14}& -S_{15}& -S_{16}& d_{11} & -d_{21}&  -d_{31}\\
& S_{22} & S_{23} & S_{24}& -S_{25}& -S_{26}& d_{12} & -d_{22}&  -d_{32}\\
& & S_{33} & S_{34}& -S_{35}& -S_{36}& d_{13} & -d_{23}&  -d_{33}\\
& & & S_{44}& -S_{45}& -S_{46}& d_{14} & -d_{24}&  -d_{34}\\
& & & & S_{55}& S_{56}& -d_{15} & d_{25}&  d_{35}\\
& & & & & S_{66}& -d_{16} & d_{26}&  d_{36}\\
& & & & & & \varepsilon_{11} & -\varepsilon_{12}&  -\varepsilon_{13}\\
& & & & & & & \varepsilon_{22}&  \varepsilon_{23}\\
& & & & & & & &  \varepsilon_{33}
\end{array}
\right), \qquad
{\bm q^s}^{\rm II} = \left(
\begin{array}{r}
{\gamma_1^s}\\
{\gamma_2^s}\\
{\gamma_3^s}\\
{\gamma_4^s}\\
-{\gamma_5^s}\\
-{\gamma_6^s}\\
{P_1^s}\\
-{P_2^s}\\
-{P_3^s}
\end{array}
\right)
\]

The domain wall is charged by discontinuous polarization components normal to the interface.
\[
\rho = \bs n \jump{\bs P^{\rm s}} = 2 P_3^{\rm s},
\]
which gives rise to internal electric fields which will ensure the continuity of dielectric displacements, even if the external field is zero.
The internal fields in both domains are directed in opposite direction to ensure a macroscopic field free state. 
The internal electric field may also result in piezoelectric strains and generate dielectric displacements parallel to the domain wall.

Obviously, the sub-matrices $Q_{ik}, i,k=1,2,6,9$ are inhomogenous for general piezoelectric materials.
Then, the simplifications for partially homogeneous laminates, made in sect. \ref{compatible-rank-1} do not apply and the general solution from sect. \ref{arbitrary_laminate} has to be used in order to compute effective properties and internal fields.

\newpage

\section{Hierarchical domain patterns (rank-two laminates)}

Ferroelectric domains are often arranged in a more complex way.
However, as compatibility of strains and dielectric displacements and energy minimization leads to essentially plane domain interfaces, the laminar structures play an prominent role. More complex patterns can be build from laminar structures by combining differently oriented laminates to so called hierarchical structures or rank-two- microstructures.  This structures can be considered as laminates of laminates, i.e. laminates, in which each lamellae again consists of a fine laminar structure.

Here we want to consider only microstructures which have  compatible spontaneous strains everywhere. 
We will denote the larger scale lamellae as {\em bands}. The interfaces between the bands are again flat.
The numerical calculation of effective properties can be performed without difficulty following the lines drawn in the preceding sections. The remaining task is to define the orientations of the individual domains and domain bands in an appropriate way. One should be aware, that the domain bands show a reduced symmetry compared to the individual domains. Therefore, we should expect even more reduced symmetries in the linear properties of the hierarchical composite.

\begin{figure}[htbp]
\begin{center}
\begin{picture}(7,7)
%\put(0,0){\framebox(7,7){}}
\put(0.2,0.5){\includegraphics[width=7cm]{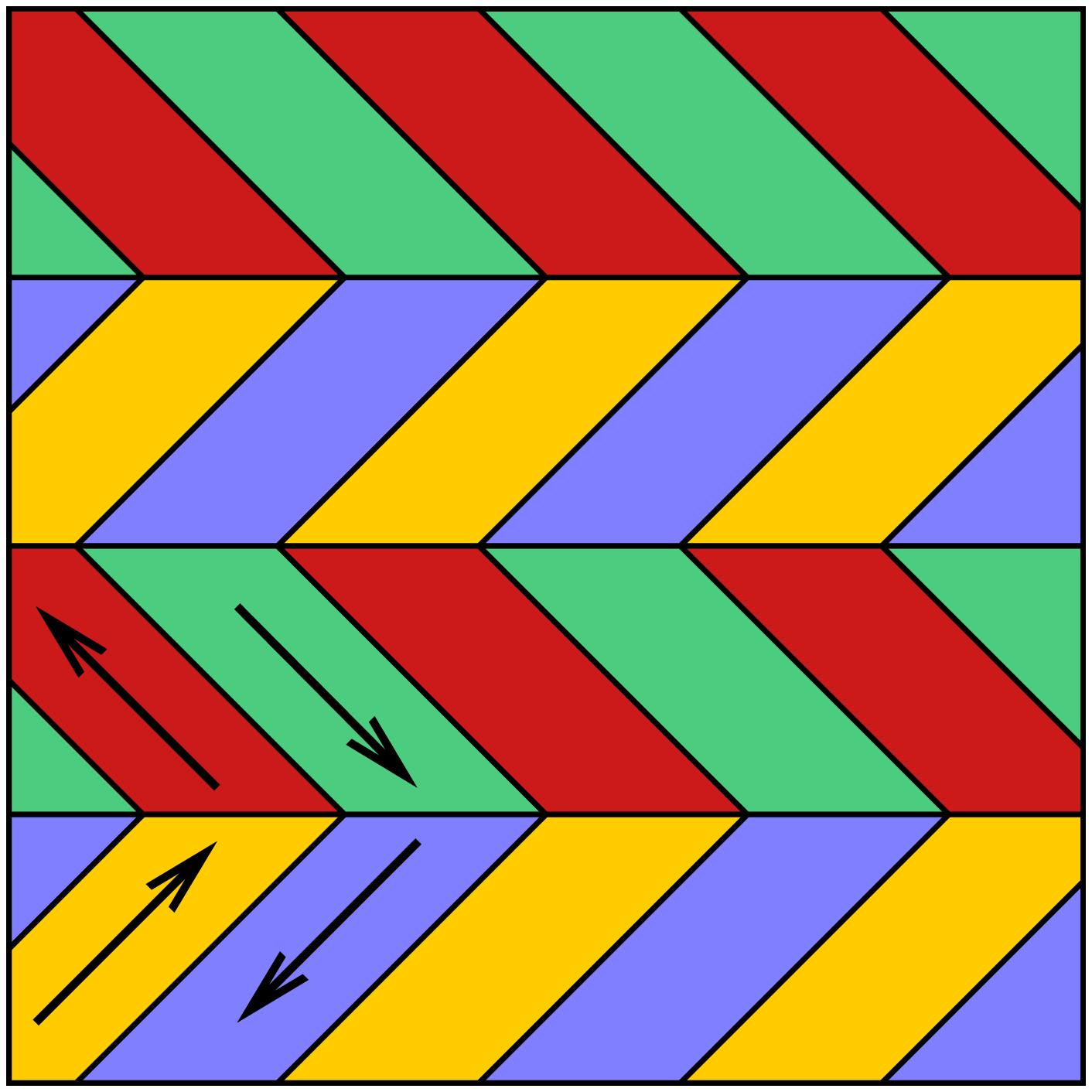}}
\put(0.,1.75){\large A $ \left\{ \rule[-5.2mm]{0pt}{13mm}\right.$}
\put(0.,3.2){\large B  $ \left\{ \rule[-5.2mm]{0pt}{13mm}\right.$}
\put(2.75,.3){\large I}
\put(1.6,.3){\large II}
\put(3.2,1.8){AI}
\put(4.25,1.8){AII}
\put(3.2,3.2){BI}
\put(4.25,3.2){BII}
\put(1.3,1.15){\large$\underbrace{\rule[0mm]{9mm}{0.mm}}$}
\put(2.4,1.15){\large$\underbrace{\rule[0mm]{9mm}{0.mm}}$}
\end{picture}
\hfill
\begin{picture}(6.5,7)
%\put(0,0){\framebox(6.5,7){}}
\put(-0.5,0.5){\includegraphics[width=7cm]{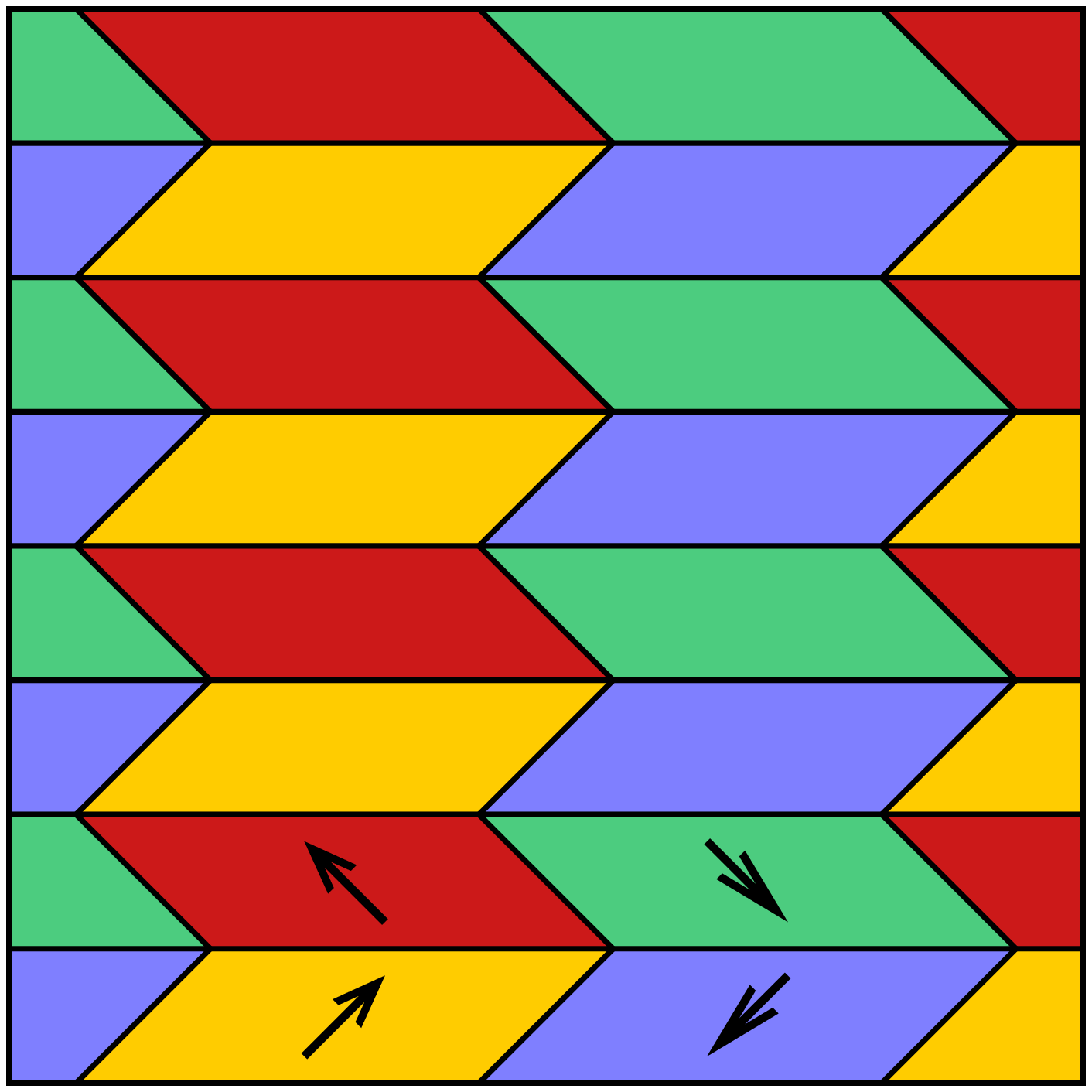}}
\put(5.8,1.4){\large $ \left. \rule[-2mm]{0pt}{6mm}\right\}$A }
\put(5.8,2.1){\large $ \left. \rule[-2mm]{0pt}{6mm}\right\}$B}
\put(2.,.3){\large I}
\put(4.,.3){\large II}
\put(1.75,2.8){AI}
\put(3.7,2.8){AII}
\put(1.75,3.5){BI}
\put(3.7,3.5){BII}
\put(0.6,1.15){\large$\underbrace{\rule[0mm]{20mm}{0.mm}}$}
\put(2.7,1.15){\large$\underbrace{\rule[0mm]{20mm}{0.mm}}$}
\end{picture}
\caption{Two equivalent examples of hierarchical domain structures }
\label{rank2domainmodels}
\end{center}
\end{figure}

We want to consider some important examples of such types of structures.
% in tetragonal ferroelectrics.
%Hierarchical rhombohedral structures are similar and should be considered elsewhere. 
The first example is a combination of tetragonal 180$^\circ$ and $90^\circ$ domains. We assume, that the fine structure is a 180$^\circ$ domain stack and two bands of 180$^\circ$ domains are arranged in such a way, that they form a 90$^\circ$ domain wall at the interface between the bands (figure \ref{rank2domainmodels}).
Furthermore, we require full compatibility of the spontaneous strain and polarization, which means that the 180$^\circ$ domain wall of the two bands meet in one point and the volume fractions of the 180$^\circ$ domains within the bands are equal.

In fact, the question which type of domains are forming the fine structure, and which combination forms the bands, is not important in our case, as there is no length scale in our model. For intrinsic linear properties of the domain structure, the only relevant additional properties are the volume fractions. Figure \ref{rank2domainmodels} shows two equivalent examples of hierarchical domain structures consisting of combined  180$^\circ$ and $90^\circ$ domains.

Starting from the 180$^\circ$ structure, described in sect. \ref{tetrag-180}, the orientation of one band is obtained by a rotation by $\approx 45^\circ$ about the $x_2$ axis. Then, the properties of the laminar structure from \ref{tetrag-180} have to be rotated using the rotation matrix 
\[
\bs \omega^A = \left(   \begin{array}{rrr}
\frac{1}{\sqrt{2}} & \pl 0 & \pl \frac{1}{\sqrt{2}}  \\
0 &1 & 0 \\
-\frac{1}{\sqrt{2}}  & 0 & \frac{1}{\sqrt{2}}
\end{array}
\right)
\]
The properties of band B are obtained applying orientation relationship accordingly \eqref{orient-relation} and with \eqref{rotation_comptbl}. Depending on the volume fractions, the average polarization varies within the square spanned by the four individual polarization vectors. The average spontaneous strain is reduced in $x_1$ and $x_3$ direction, but remains unchanged in perpendicular direction. If a crystal is embedded within a elastic neighborhood, hierarchical microstructures can reduce the internal stresses and the internally stored energy \citep{arlt.90}. In this case, such an in-plane arrangement of domains does not permit an optimal relaxation of internal energy \citep{arlt.80}.
 
\begin{figure}[htbp]
\unitlength 1. cm
\begin{center}
\begin{picture}(10,11.5)
%\put(0,0){\framebox(10,11.5){}}
\put(0,0){\includegraphics[width=10cm]{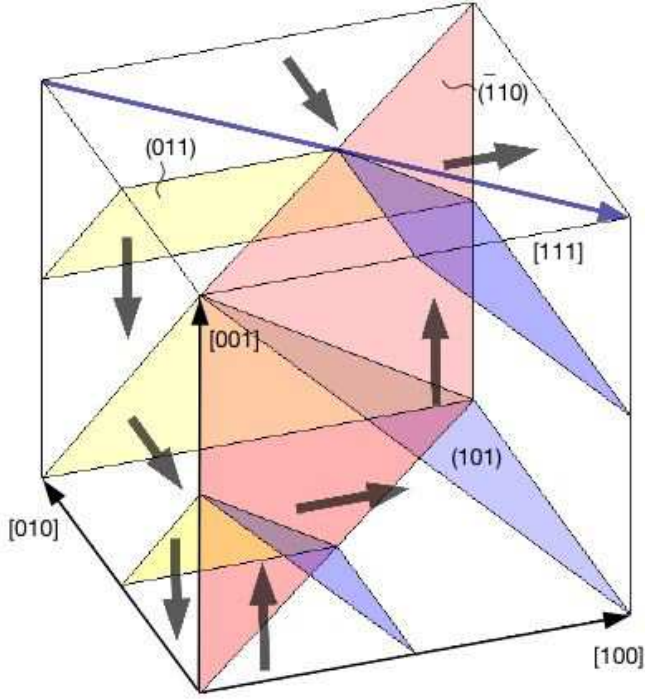}}
\end{picture}
\caption{Three dimensional arrangement of four different polarization directions,  with effective polarization along pseudo-cubic [1$\bar 1$0], \citep{arlt.80}}
\label{arlts_cube}
\end{center}
\end{figure}

A domain arrangement, which is often observed in ceramic materials, may be characterized by a 3-dimensional arrangement of 
ferroelectric domains (see fig. \ref{arlts_cube}). Two bands of 90$^\circ$ domain structures ([001]/[100] with (101) interface and 
$[00 \bar 1]$/$[0 \bar 1 0]$ with (011) interface) are combined in such a way, that the interface between the bands is the pseudo-cubic $(\bar 1 1 0)$ plane. Two domains form a 90$^\circ$ domain structure across the band interface  ([001]/$[00 \bar 1]$) and two domains form a 90$^\circ$ domain structure ([100] /$[0 \bar 1 0]$). Assuming equal volume fraction for all four domain, then the first domain band has an average polarization along [101] and the second band along $[0 \bar 1 \bar 1]$ resulting in an average polarization of the hierarchical structure $\bs P^{\rm s\, \ast} = (1, -1, 0) P^{\rm s}/4$.

Starting from the 90$^\circ$ domain structure from sect. \ref{tetrag-90}, a series of three rotations is necessary to transform the crystal into a suitable orientation with the interface normal of the $(\bar 1 1 0)$ plane parallel to the macroscopic $x_3$ direction:
a first rotation by 45$^\circ$ about $x_2$ brings the crystal back into an orientation with the pseudo-cubic axes parallel to the axes of the  macroscopic coordinate system. A rotation by -45$^\circ$ about $x_3$ and by 90$^\circ$ about $x_1$ gives the final orientation of the band A with
\[
\bs \omega^A = \left(   \begin{array}{ccc}
\;\frac{1}{2} & \frac{1}{\sqrt{2}}  & \pl \frac{1}{2}   \\
\frac{1}{\sqrt{2}}  & 0 & -\frac{1}{\sqrt{2}}  \\
-\frac{1}{2}\pl  & \frac{1}{\sqrt{2}}  & -\frac{1}{2} 
\end{array}
\right)
\]
The properties of band B are obtained applying orientation relationship accordingly \eqref{orient-relation} and with \eqref{rotation_comptbl}.

\begin{figure}[htbp]
\unitlength 1. cm
\begin{center}
\begin{picture}(10,10)
\put(0,0){\includegraphics[width=10cm]{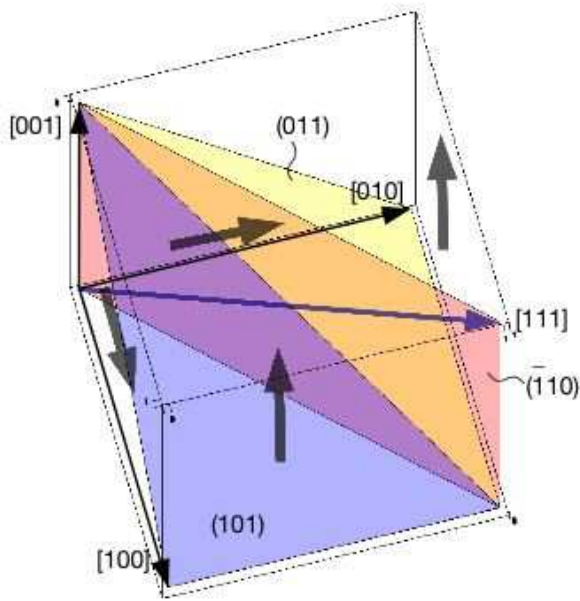}}
%\put(0,0){\framebox(10,10){}}
\end{picture}
\caption{Arrangement of three different polarization directions in a tetragonal ferroelectric, which form an engineered domain structure with polarization along pseudo-cubic [111]}
\label{cube_engineered_111}
\end{center}
\end{figure}

A closely related example is that of a engineered domain structure in a tetragonal ferroelectric, poled along (111) (fig. \ref{cube_engineered_111}).
The three orientations [100], [010], and [001] are arranged in two bands forming a mechanically compatible domain structure, but with partially incompatible polarizations on the ($\bar 1$10) interface between the band. Band A and its orientation in the macroscopic coordinate system is identical to that of the preceding example. Band B consists of  [010] and [001] domains with the domain wall (011). 
The orientation of band B in the macroscopic coordinate system is then obtained by applying a rotation by -90$^\circ$ about $x_2$.
That part of the interface, where the [100]-domains of Band and the [010]-domains of band B meet, is charged.
If the volume fraction of the [001] domains in the two bands is $1/3$ and the volume fractions of the two bands are equal, then the overall polarization is along [111].

\begin{figure}[htbp]
\unitlength 1. cm
\begin{center}
\begin{picture}(10,10)
\put(1,1){\includegraphics[width=8cm]{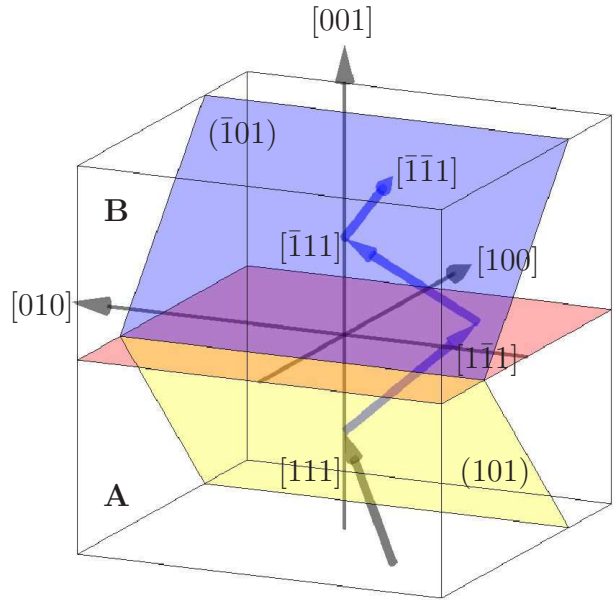}}
%\put(0,0){\framebox(10,10){}}
\put(5,9){\makebox(0,0){[001]}}
\put(1,5.2){\makebox(0,0){[010]}}
\put(7.2,5.8){\makebox(0,0){[100]}}
\put(5,3){\makebox(0,0)[r]{[111]}}
\put(5.7,7){\makebox(0,0)[l]{$[\bar 1 \bar 11]$}}
\put(6.5,4.5){\makebox(0,0)[l]{$[1 \bar 1 1]$}}
\put(5,6){\makebox(0,0)[r]{$[\bar 111]$}}
\put(3.2,7.5){\makebox(0,0)[l]{$(\bar 101)$}}
\put(7.5,3){\makebox(0,0)[r]{$(101)$}}
\put(2,2.7){\makebox(0,0){\bf A}}
\put(2,6.5){\makebox(0,0){\bf B}}
\end{picture}
\caption{Rhombohedral domain structure with average polarization along pseudo-cubic [001]}
\label{rh-chevron-001}
\end{center}
\end{figure}

A final example demonstrates a hierarchical domain structure for rhombohedral crystals poled in the  pseudo-cubic [001] direction (fig. \ref{rh-chevron-001}).
It consists of two domain bands with 109$^\circ$ domain structures, [111]/$[1 \bar 1 1]$ with (101) domain wall and $[\bar 1 11]$/$[\bar 1 \bar 11]$ with $(\bar 101)$ domain walls. The interface between the bands is the (001) plane. Within each band, the polarizations lie with the $(\bar 101)$  and (101) plane, respectively. For analysis of the effective properties of this structure, it is suitable to depart from the 109$^\circ$ domain structure in sect. \ref{rh-domains}. This has to be rotated by 90$^\circ$ about $x_3$ and by 45$^\circ$ about $x_2$ in order to arrive at the orientation of the lower half of figure \ref{rh-chevron-001}. The rotation matrix, which transforms the 109$^\circ$ laminate into band A is
\[
\bs \omega^A = \left(   \begin{array}{crc}
\frac{1}{\sqrt{2}} & \pl 0 & \pl\frac{1}{\sqrt{2}}   \\
\frac{1}{\sqrt{2}} & 0 & -\frac{1}{\sqrt{2}}  \\
0  & 1 & \pl 0
\end{array}
\right)
\]
The upper half, band B, is again obtained by the orientation relationship of a fully compatible structure, \eqref{orient-relation} and \eqref{rotation_comptbl}. Across the band interface, the domains form 71$^\circ$ domain structures with polarizations in the ($1 \bar 1 0$) and (110) plane, respectively. The average polarization of the composite structure is along [001], provided that all domains have the same volume fraction.

\newpage

\section{Numerical results}

\begin{figure}[htbp]
\begin{center}
\includegraphics[width=12cm]{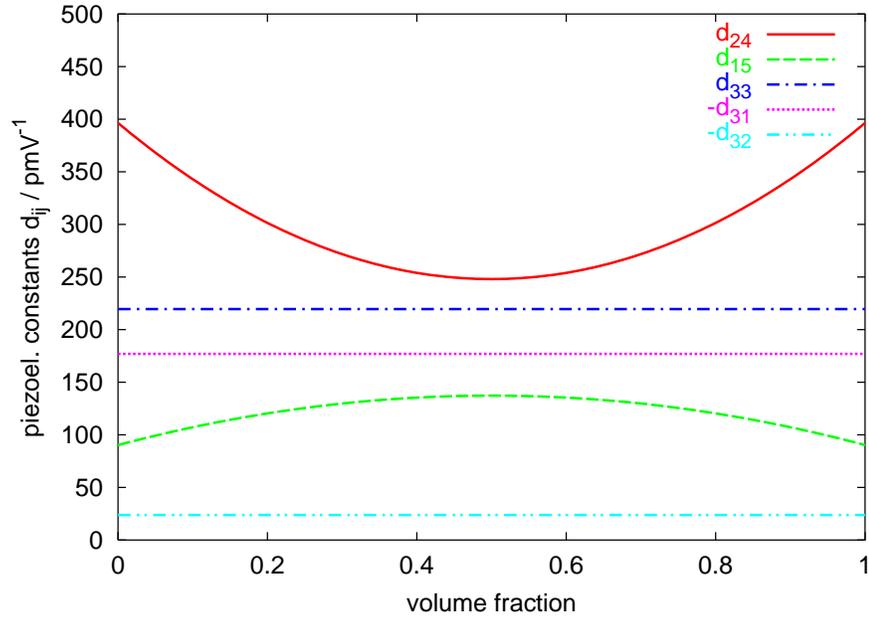}
\caption{Piezoelectric coefficients for a 90$^\circ$ domain structure of  tetragonal $\rm BaTiO_3$, dependent upon the volume fraction of domain I $\xi$}
\label{stackpiezo_vol1}
\end{center}
\end{figure}

The theory presented in this paper has been used to evaluate the effective properties for various arrangements of ferroelectric domains in tetragonal Barium Titanate. This material is chosen as it is well characterized and may serve as a prototype material for tetragonal ferroelectrics.
The results for distinct cases are compiled in table \ref{numbers}. 
The single crystal properties are taken from \citep{zgonik.94} and ere described in an orientation where
$[001]\; || x_3$. 
The case of 90$^\circ$ laminate refers to an simple laminar domain structure with $\bs n \; || [101]\; || x_3$ and a volume fraction 50\% for each domain type. 
Additionally, the bounds on the effective properties, $\averg{\bm Q}$ and $\averg{\bm R}$, are given. The dependence of the piezoelectric coefficients $d_{31}^\ast$, $d_{32}^\ast$, $d_{33}^\ast$, $d_{24}^\ast$ and $d_{15}^\ast$ on the volume fraction is plotted in fig. \ref{stackpiezo_vol1}. In this case, the piezoelectric constants $d_{31}^\ast$ and $d_{33}^\ast$ increase by more than 400\% and 100\%, respectively, compared to the single crystal values and are independent upon the volume fraction.
$d_{32}^\ast$ is slightly smaller than the single crystal value. 
These effects are, simple averaging effects, i.e. the increase of the $d_{31}^\ast$ and $d_{33}^\ast$ are only a result of the rotation of the crystal and thus caused by the contribution of the large shear coefficient $d_{15}$.

\begin{figure}[htbp]
\begin{center}
\includegraphics[width=12cm]{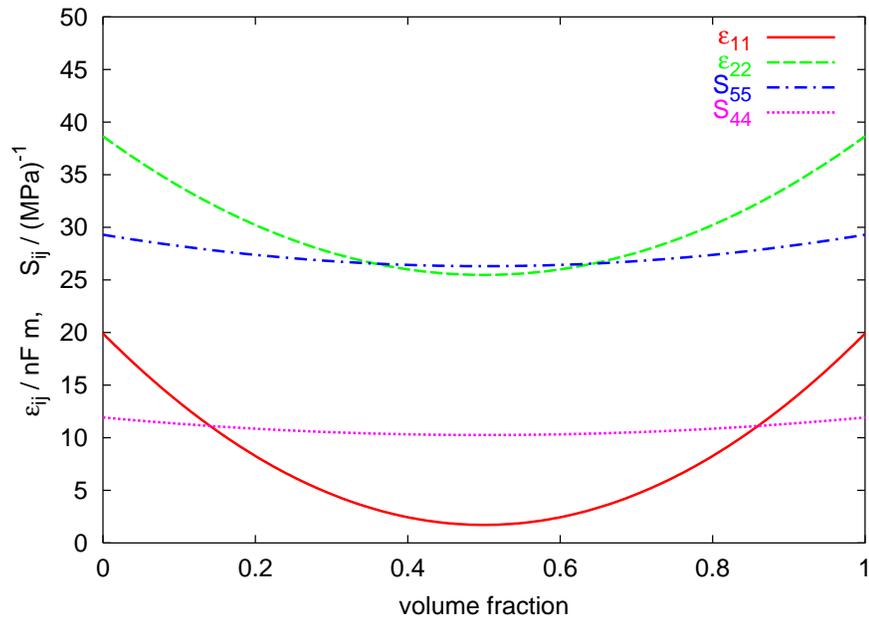}
\caption{Dielectric and elastic coefficients,  $S_{44}^\ast$, $S_{55}^\ast$, $\varepsilon_{11}^\ast$ and $\varepsilon_{22}^\ast$  for a 90$^\circ$ domain structure of  tetragonal $\rm BaTiO_3$, dependent upon the volume fraction of domain I }
\label{stack44-55-77-88}
\end{center}
\end{figure}

More interesting are the piezoelectric coefficients $d_{24}^\ast$ and $d_{15}^\ast$, as they show a dependence on the interaction between the domains, i.e. these effective properties differ from the lower bound (Reuss approximation).
They show a quadratic dependency upon the volume fraction with a minimum or maximum, respectively, at $\xi = 0.5$. The effective $d_{15}^\ast$ is slightly boosted by domain interaction, but $d_{24}^\ast$ is strongly restrained.
The elastic constants $S_{44}^\ast$, and $S_{55}^\ast$, as well as the dielectric constants $\varepsilon_{11}^\ast$ and $\varepsilon_{22}^\ast$  are plotted in fig. \ref{stack44-55-77-88}, which shows, that the elastic coefficients are much less effected by domain interactions compared with the dielectric coefficients.

The overall symmetry of a simple laminar domain structure is clearly orthorhombic (point group {\em mm2}) with special relationships $S_{12}=S_{23}$ and $S_{11}=S_{33}$. Obviously, the shear components $S_{44}$, $S_{55}$ and $S_{66}$ are all different, which is in contrast to the finding of  \citep{erhart.99}.

Comparing the effective properties with the upper and lower bounds, one has to note, that a particular component not necessarily  has to be between the values of the bounds, but the differences between the property matrices have to be positive semi-definite, which is true in the studied cases.

Furthermore, the effective properties of a 90$^\circ $ domain structure with charged domain walls are calculated.The results obtained from the general formula in sect. \ref{arbitrary_laminate} are calculated with a domain wall normal $\bs n || x_3$ and the symmetry relationship \eqref{rotation_charged} and a volume fraction $\xi=0.5$.
The average polarization is parallel to the domain wall. For sake of comparison we have finally rotated the results so that the average polarization is in $x_3$ direction, i.e. the interface normal is $\bs n = (1,0,0)^t$. This configuration is characterized by much decreased piezoelectric constants, except $d_{32}^\ast$. It is remarkable, that the sign of $d_{31}^\ast$ is changed.
In table \ref{numbers-strain-pol} the effective strains and polarization of the two different $90^\circ$ domain configurations are compared with the single crystal values. It demonstrates, how much the effective polarization is increased due to the internal electric fields in the case of a charged domain wall.

Finally, a comparison of the effective properties of three different hierarchical domain structures in tetragonal $\rm BaTiO_3$ is given in table \ref{numbers2}. The properties are given with respect to the coordinate system, in which they are calculated, i.e. the macroscopic $x_3$ axis is perpendicular to the band interfaces. In the first case of a $180^\circ/90^\circ$ domain strcuture with equal volume fractions, the macroscopic piezoelectric effect vanishes. Elastic and dielectric properties have othorhombic symmetry {\em mm2}.
In the following cases of a fully compatible 3-dimensional arrangement of 90$^\circ$ domains according to \citet{arlt.80} and the charged 3-dimensional arrangement of 90$^\circ$ domains, the over-all symmetry is not further investigated. It is however obvious, that the symmetry is strongly reduced. 
In the case of the charged arrangement, we have calculated a polarization of $\bs P^{\rm s} = (0.2,  -0.04,0)^t$, which is not the pseudo-cubic [111] direction. This effect is again caused by the large internal electric fields, which amplify the dielectric displacements in the macroscopic $x_1$ direction.

\begin{table}[hbtp]
\caption{Linear material properties for barium titanate in different domain configurations: Single domain single crystals (s. cryst.), uncoupled domains I and II (rotated single crystal), compatible 90$^\circ$ domain structure and a 90$^\circ$ domain structure with charged domain wall, both for a volume fraction $\xi = 0.5$. 
Elastic constants are the stiffness at zero electric field in $\rm 10^{-3}\,GPa^{-1}$, the piezoelectric constants are given in $\rm pm V^{-1}$ and the dielectric constants are at zero stress, given in $\rm 10^{-9}\, Fm$ }
\renewcommand{\arraystretch}{1.2}
\setlength{\tabcolsep}{2ex}
\begin{center}
\begin{tabular}[t]{|c|r| r | r | r | r | r|}
\hline

& s. cryst. & I, II &  compatible & charged & \multicolumn{2}{c|}{bounds \rule{0pt}{2.5ex}}\\[1ex]\cline{6-7}
& $\uparrow$ & $\nearrow , \nwarrow$ & $\displaystyle \frac{\nwarrow}{\nearrow}$ &  
$\displaystyle  \nearrow | \nwarrow$ 
&$\averg{\bm Q}$ & $\averg{\bm R}^{-1}$\\[2ex]
\hline
$S_{11}$ &  7.383 &       7.016  &  7.016 &  4.596 &  7.016 &  4.476 \\
$S_{12}$ & -1.389 &      -2.898  & -2.898 & -2.635 & -2.898 & -2.447 \\
$S_{13}$ & -4.407 &      -1.181  & -1.181 &  0.770 & -1.181 &  0.534 \\
%$S_{14}$ &  0     &       0      &  0     &  0     &  0     &  0     \\
$S_{15}$ &  0     &$\pm$  2.859  &  0     &  0     &  0     &  0     \\
%$S_{16}$ &  0     &       0      &  0     &  0     &  0     &  0     \\
$S_{22}$ &  7.383 &       7.383  &  7.383 &  7.355 &  7.383 &  7.062 \\
$S_{23}$ & -4.407 &      -2.898  & -2.898 & -3.110 & -2.898 & -2.741 \\
%$S_{24}$ &  0     &       0      &  0     &  0     &  0     &  0     \\
$S_{25}$ &  0     &$\mp$  3.018  &  0     &  0     &  0     &  0     \\
%$S_{26}$ &  0     &       0      &  0     &  0     &  0     &  0     \\
$S_{33}$ & 13.101 &       7.016  &  7.016 &  5.444 &  7.016 &  4.980 \\
%$S_{34}$ &  0     &       0      &  0     &  0     &  0     &  0     \\
$S_{35}$ &  0     &$\pm$  2.859  &  0     &  0     &  0     &  0     \\
%$S_{36}$ &  0     &       0      &  0     &  0     &  0     &  0     \\
$S_{44}$ & 16.393 &      11.928  & 10.256 & 11.928 & 11.928 & 10.256 \\
%$S_{45}$ &  0     &       0      &  0     &  0     &  0     &  0     \\
$S_{46}$ &  0     &$\pm$  4.465  &  0     &  0     &  0     &  0     \\
$S_{55}$ & 16.393 &      29.299  & 26.307 & 27.588 & 29.299 & 26.018 \\
%$S_{56}$ &  0     &       0      &  0     &  0     &  0     &  0     \\
$S_{66}$ &  7.463 &      11.928  & 11.928 & 10.256 & 11.928 &  7.842 \\
\hline
%\end{tabular}
%\begin{tabular}[t]{|c|r|}
%\hline
%& single crystal\\
%\hline
$\varepsilon_{11}$ & 38.653 &      19.913 &  1.714 & 13.933 & 19.913 &  1.644 \\
%$\varepsilon_{12}$ &  0     &       0     &  0     &  0     &  0     &  0     \\
$\varepsilon_{13}$ &  0     &$\mp$ 18.740 &  0     &  0     &  0     &  0     \\
$\varepsilon_{22}$ & 38.653 &      38.653 & 25.477 & 38.653 & 38.653 & 25.477 \\
%$\varepsilon_{23}$ &  0     &       0     &  0     &  0     &  0     &  0     \\
$\varepsilon_{33}$ & 1.172  &      19.913 & 19.913 &  2.276 & 19.913 &  1.213 \\
\hline
$d_{11}$ &   0     &$\pm$  219.515 &    0     &   0     &    0     &   0    \\
$d_{12}$ &   0     &$\mp$   23.845 &    0     &   0     &    0     &   0    \\
$d_{13}$ &   0     &$\mp$  176.928 &    0     &   0     &    0     &   0    \\
%$d_{14}$ &   0     &         0     &    0     &   0     &    0     &   0    \\
$d_{15}$ & 560.656 &        90.278 &  137.242 & 133.588 &   90.278 & 132.760\\
%$d_{16}$ &   0     &         0     &    0     &   0     &    0     &   0    \\
%$d_{21}$ &   0     &         0     &    0     &   0     &    0     &   0    \\
%$d_{22}$ &   0     &         0     &    0     &   0     &    0     &   0    \\
%$d_{23}$ &   0     &         0     &    0     &   0     &    0     &   0    \\
$d_{24}$ & 560.656 &       396.443 &  248.031 & 396.443 &  396.443 & 248.031\\
%$d_{25}$ &   0     &         0     &    0     &   0     &    0     &   0    \\
$d_{26}$ &   0     &$\pm$  396.443 &    0     &   0     &    0     &   0    \\
$d_{31}$ & -33.722 &      -176.928 & -176.928 &  29.660 & -176.928 &  18.354\\
$d_{32}$ & -33.722 &       -23.845 &  -23.845 & -46.286 &  -23.845 & -28.633\\
$d_{33}$ &  93.949 &       219.515 &  219.515 &  53.006 &  219.515 &  30.798\\
%$d_{34}$ &   0     &         0     &    0     &   0     &    0     &   0    \\
$d_{35}$ &   0     &$\pm$   90.278 &    0     &   0     &    0     &   0    \\
%$d_{36}$ &   0     &         0     &    0     &   0     &    0     &   0    \\
\hline
\end{tabular}
\end{center}
\label{numbers}
\end{table}

\newpage

\begin{table}[hbtp]

\caption{Spontaneous strains and polarizations of tetragonal $\rm BaTiO_3$ single crystals, in the reference orientation and rotated, and effective strains and polarization of fully compatible and charged 90$^\circ$ domain structures. Volume fraction $\xi = 0.5$. Polarizations are given in $\rm C/m^2$}

\setlength{\tabcolsep}{2ex}
\begin{center}
\begin{tabular}[t]{|c|r| r | r | r |}
\hline
& s. cryst. & I, II &  compatible & charged \\[1ex]
& $\uparrow$ & $\nearrow , \nwarrow$ & $\displaystyle \frac{\nwarrow}{\nearrow}$ &  
$\displaystyle  \nearrow | \nwarrow$ \\
\hline
$\gamma_{11}^{\rm s}$ & -0.0035 &      0.00175 &   0.00175 & -0.00023 \\
$\gamma_{22}^{\rm s}$ & -0.0035 &     -0.00350 &  -0.00350 & -0.00327 \\
$\gamma_{33}^{\rm s}$ &  0.0070 &      0.00175 &   0.00175 &  0.00338 \\
% $\gamma_{23}^{\rm s}$ &  0.0000 &      0.00000 &   0.00000 &  0.00000 \\
$\gamma_{13}^{\rm s}$ &  0.0000 & $\pm$0.01050 &   0.00000 &  0.00000 \\
% $\gamma_{12}^{\rm s}$ &  0.0000 &      0.00000 &   0.00000 &  0.00000 \\
\hline
      $P_{1}^{\rm s} $ &  0.0000 & $\pm$0.18385 &   0.00000 &  0.00000 \\
%       $P_{2}^{\rm s}$ &  0.0000 &      0.00000 &   0.00000 &  0.00000 \\
      $P_{3}^{\rm s}$ &  0.2600 &      0.18385 &   0.18385 &  0.35590 \\
\hline
\end{tabular}
\end{center}
\label{numbers-strain-pol}
\end{table}

\begin{table}[hbtp]
\caption{Linear material properties for barium titanate in different hierarchical domain configurations: 
Units as in tab. \ref{numbers} }
\renewcommand{\arraystretch}{1.2}
\setlength{\tabcolsep}{2ex}
\begin{center}
\begin{tabular}[t]{|c| r | r | r |}
\hline
&$180^\circ /90^\circ$ & compatible $90^\circ /90^\circ$& charged  $90^\circ /90^\circ$\\
& $\displaystyle \frac{\nwarrow \searrow}{\nearrow \swarrow}$ 
& $\displaystyle \searrow \downarrow \uparrow \nearrow $
& $\displaystyle \nwarrow \uparrow \uparrow \nearrow $ \\

\hline
$S_{11}$           &   4.983 &   5.544 &   5.555 \\
$S_{12}$           &  -2.898 &  -3.278 &  -3.197 \\
$S_{13}$           &   0.852 &  -0.002 &   0.192 \\
% $S_{14}$           &   0     &   0 &   0 \\
% $S_{15}$           &   0     &   0 &   0 \\
$S_{16}$           &   0     &   0.296 &   0.034 \\
$S_{22}$           &   7.383 &   9.495 &   8.580 \\
$S_{23}$           &  -2.898 &  -3.278 &  -2.876 \\
% $S_{24}$           &   0     &   0 &   0 \\
% $S_{25}$           &   0     &   0 &   0 \\
$S_{26}$           &   0     &   0     &  -0.076 \\
$S_{33}$           &   4.983 &   5.544 &   5.081 \\
% $S_{34}$           &   0     &   0 &   0 \\
% $S_{35}$           &   0     &   0 &   0 \\
$S_{36}$           &   0     &  -0.296 &   0.078 \\
$S_{44}$           &  10.256 &  10.350 &  12.161 \\
$S_{45}$           &   0     &   0.892 &   0.074 \\
% $S_{46}$           &   0     &   0 &   0 \\
$S_{55}$           &  27.336 &  22.229 &  24.262 \\
% $S_{56}$           &   0     &   0 &   0 \\
$S_{66}$           &  11.928 &  13.743 &  12.910 \\
\hline 
$\varepsilon_{11}$ &   0.983 &   9.881 &   9.231 \\
$\varepsilon_{12}$ &   0     &  -5.783 &  -9.474 \\
% $\varepsilon_{13}$ &   0     &   0 &   0 \\
$\varepsilon_{22}$ &  19.479 &   5.546 &  16.078 \\
% $\varepsilon_{23}$ &   0     &   0 &   0 \\
$\varepsilon_{33}$ &  19.575 &  18.145 &  14.405 \\
\hline
$d_{11}$           &   0     &   0 &   66.184 \\
$d_{12}$           &   0     &   0 &  -43.650 \\
$d_{13}$           &   0     &   0 &  -1.814 \\
$d_{14}$           &   0     &  97.187 &   0 \\
$d_{15}$           &   0     & -68.931 &   0 \\
$d_{16}$           &   0     &   0 &  -170.729 \\
$d_{21}$           &   0     &   0 &  -9.464 \\
$d_{22}$           &   0     &   0 &  -55.297 \\
$d_{23}$           &   0     &   0 &   60.330 \\
$d_{24}$           &   0     &  84.702 &   0 \\
$d_{25}$           &   0     &  33.530 &   0 \\
$d_{26}$           &   0     &   0 &   248.037 \\
$d_{31}$           &   0     &  45.491 &   0 \\
$d_{32}$           &   0     &  23.664 &   0 \\
$d_{33}$           &   0     & -78.525 &   0 \\
$d_{34}$           &   0     &   0 &  -134.113 \\
$d_{35}$           &   0     &   0 &   86.450 \\
$d_{36}$           &   0     &  52.472 &   0 \\
\hline
\end{tabular}
\end{center}
\label{numbers2}
\end{table}

\clearpage

\newpage

\section{Summary}

The effective properties of piezoelectric laminates have been analyzed. The analysis is based on the calculation of internal fields and makes use of a simple matrix manipulation method. The results are expressed in a compact notation which is convenient for numerical implementation and at the same time 
suitable for further analytical treatments.

A more detailed analysis of fully compatible ferroelectric domain structures shows, that the results for arbitrary piezoelectric laminates can be further simplified and specific property relationships for rank-1 laminates of tetragonal and rhombohedral crystals are derived. In particular, it is shown that only the components $S_{44}$, $S_{45}$ and $S_{55}$ as well as $\varepsilon_{11}$, $\varepsilon_{12}$ and $\varepsilon_{22}$ are effected by interactions between the different domains. All other components can be also obtained by simple volume averaging of the single domain properties, i.e. they are equal to the Reuss lower bound on the effective properties.

In contrast, in the case of charged domain structure, where the spontaneous strains are compatible but not the polarizations, the conditions for a simplified treatment are not fulfilled. 

The method is finally applied to the analysis of various hierarchical domain structures (rank-2 laminates, or laminates of laminates). Detailed orientation relationships between the particular domains in some important domain pattern are given to make these structures accessible for the presented method.
Some numerical results for barium titanate are given in order to illustrate the effects of different domain arrangements on the effective properties.

%\newpage

\section*{Acknowledgment}

The financial support of this work by the Deutsche Forschungsgemeinschaft under grant Ro 2420/1 is gratefully acknowledged.
The author is particular grateful to W. S. Kreher, Technische Universit\"at Dresden, who originally provided the solution for arbitrary piezoelectric laminates. 
Furthermore, thanks goes to P. Obenaus and and T. Scholehwar, Fraunhofer IKTS Dresden, for support with SEM imaging of domain structures.

\clearpage

 \appendix

\section{Detailed solution of the internal fields for the arbitrary laminate}
\label{detail_A}

A detailed solution of the internal fields \eqref{local_field_p} with the conditions (\ref{stackcond_q}-\ref{stackcond_p2}) for the arbitrary laminate is demonstrated in the following. 
The solution is analogous to that one derived by \citet{report.MW282710}, but stress and electric field are chosen to be independent variables.

Although we have given the result for the interaction matrices and vectors in eqns. (\ref{sum_arbsol_a}-\ref{sum_arbsol_e}) in an index notation, we will derive the general solution in this appendix in a symbolic way, which enables a very lucid derivation of the results and is convenient for numerical implementation.  
This representation of the problem is equivalent to that of \citet{liu-li.03} and \citet{li.04}. However, their derivation of the results is slightly different.
The results can easily be converted in the usual notation, simply by rearranging the vector and matrix component in an appropriate way.

We rearrange the components of the vectors $\bm p$ and $\bm q$ , so that the first four components represent the field components, for which the strain and dielectric displacement are continuous across the interface.
For the remaining five components, the stress and electric field is continuous. 
%\[
%\bm q \to \left( \begin{array}{c}
% q_1 \\ q_2 \\ q_6 \\ q_9  \\  q_3 \\ q_4 \\ q_5 \\ q_7 \\ q_8
%\end{array} \right)
%= \left(  \begin{array}{l}  
%             \bm q^a  \\
%             \bm q^b 
% \end{array} \right), 
% = \left(  \begin{array}{l}  
%             q_{\tilde i}  \\
%             q_{\tilde {\tilde i}}
% \end{array} \right), 
% \qquad
%\bm p \to \left( \begin{array}{c}
% p_1 \\ p_2 \\ p_6 \\ p_9  \\  p_3 \\ p_4 \\ p_5 \\ p_7 \\ p_8
%\end{array} \right)
%= \left(  \begin{array}{l}  
%              \bm p^a  \\
%             \bm p^b 
%              \end{array} \right),
%= \left(  \begin{array}{l}  
%             p_{\tilde i}  \\
%             p_{\tilde {\tilde i}}
% \end{array} \right)\]
\[
\bm q^a = \left( \begin{array}{c}  q_1 \\ q_2 \\ q_6 \\ q_9  \end{array} \right) , \quad 
\bm q^b = \left( \begin{array}{c}q_3 \\ q_4 \\ q_5 \\ q_7 \\ q_8 \end{array} \right), 
\qquad
\bm p^a = \left( \begin{array}{c}p_1 \\ p_2 \\ p_6 \\ p_9  \end{array} \right),  \quad
\bm p^b = \left( \begin{array}{c}p_3 \\ p_4 \\ p_5 \\ p_7 \\ p_8 \end{array} \right)              
\]
Then, the constitutive law takes the form:
\begin{equation}\label{const_eq_ab}
 \left(  \begin{array}{l}  
             \bm q^a   \\
             \bm q^b
 \end{array} \right)
 =  \left(  \begin{array}{cc}  
             \bm Q^{aa} & \bm Q^{ab}   \\
             \bm Q^{ba} & \bm Q^{bb}
 \end{array} \right)
 \left(  \begin{array}{l}  
             \bm p^a   \\
             \bm p^b
 \end{array} \right)
 + \left(  \begin{array}{l}  
             {\bm q^{\rm s}}^a   \\
             {\bm q^{\rm s}}^b
 \end{array} \right)
\end{equation}
with
\[
\bm Q^{aa} = \left(  \begin{array}{cccc}  
             Q_{11} & Q_{12} & Q_{16} & Q_{19}  \\
             Q_{12} & Q_{22} & Q_{26} & Q_{29}  \\
             Q_{16} & Q_{26} & Q_{66} & Q_{69}  \\
             Q_{19} & Q_{29} & Q_{69} & Q_{99}  
 \end{array} \right),\qquad
\bm Q^{ab} = \left(  \begin{array}{ccccc}  
             Q_{13} & Q_{14} & Q_{15} & Q_{17} & Q_{18} \\
             Q_{23} & Q_{24} & Q_{25} & Q_{27} & Q_{28} \\
             Q_{36} & Q_{46} & Q_{56} & Q_{67} & Q_{68} \\
             Q_{39} & Q_{49} & Q_{59} & Q_{79} & Q_{89}  
\end{array} \right)
\]
\[
\bm Q^{ba} = {\bm Q^{ab}}^t, 
\qquad
\bm Q^{ab} = \left(  \begin{array}{ccccc}  
             Q_{33} & Q_{34} & Q_{35} & Q_{37} & Q_{38} \\
             Q_{34} & Q_{44} & Q_{45} & Q_{47} & Q_{48} \\
             Q_{35} & Q_{45} & Q_{55} & Q_{57} & Q_{58} \\
             Q_{37} & Q_{34} & Q_{57} & Q_{77} & Q_{78} \\
             Q_{38} & Q_{48} & Q_{58} & Q_{78} & Q_{88} \\
\end{array} \right)
\]
Correspondingly, the local fields $\bm p^{(k)}$ are expressed by
\begin{equation}\label{A2}
 \left(  \begin{array}{l}  
             \bm p^a   \\
             \bm p^b
 \end{array} \right)
 =  \left(  \begin{array}{cc}  
             \bm A^{aa} & \bm A^{ab}   \\
             \bm A^{ba} & \bm A^{bb}
 \end{array} \right)
 \left(  \begin{array}{l}  
             \bar {\bm p}^a   \\
             \bar {\bm p}^b
 \end{array} \right)
 + \left(  \begin{array}{l}  
             {\bm p^a}^{A}   \\
             {\bm p^b}^{A}
 \end{array} \right)
\end{equation}
The conditions (\ref{stackcond_q} -\ref{stackcond_p2}) are
{\setcounter{eqnstop}{\arabic{equation}}
\addtocounter{eqnstop}{1} % bisher steht da der letzte Wert drin!
\setcounter{equation}{0}
\renewcommand{\theequation}{\Alph{section}.\arabic{eqnstop}.\alph{equation}}
\begin{align}
{\bm q^a}^{(k)} &= {\bm q^a}^{(1)} 
\label{stackcond_qa}
\\
{\bm p^b}^{(k)} &= {\bm {\bar p}^b} & 
\label{stackcond_pb}
\\
\averg{ {\bm p^a}^{(k)}} &= \bm {\bar p}^a  
\label{stackcond_pa}
\end{align}
\setcounter{equation}{\arabic{eqnstop}}
} 
\eqref{stackcond_p1} or \eqref{stackcond_pb} yields directly:
\begin{equation}
{\bm A^{bb}}^{(k)} = \bm I^{bb} , \qquad 
{\bm A^{ba}}^{(k)} = \bm 0 , \qquad 
{\bm p^b}^{\rm A\, (k)} = 0  \nonumber
\end{equation}
where $\bm I^{bb} = \delta_{\rm ij}$, $\rm i,j = 1...5$.
This results are equivalent to \eqref{sum_arbsol_a} and  \eqref{sum_arbsol_d}.

\eqref{stackcond_qa} can be written in the following way, 
using \eqref{const_eq_ab}
\begin{equation}
{\bm q^a}^{(k)} = {\bm Q^{aa}}^{(k)} {\bm p^a}^{(k)} + {\bm Q^{ab}}^{(k)} {\bm p^b}^{(k)} 
+ {{\bm q^{\rm s}}^a}^{(k)}
= {\bm q^a}^{(1)}
\end{equation}
With the abbreviation ${\bm a }^{(k)} =  {{\bm q^{\rm s}}^a}^{(k)} + {\bm Q^{ab}}^{(k)} {\bm p^b}^{(k)} $
%and \eqref{stackcond_pb} 
we get
\begin{equation}
{\bm Q^{aa}}^{(k)} {\bm p^a}^{(k)}
= {\bm q^a}^{(1)} - \bm a^{(k)}
\end{equation}
\begin{equation} \label{A5}
 {\bm p^a}^{(k)}
=\left( {\bm Q^{aa}}^{(k)} \right)^{-1} \left(   {\bm q^a}^{(1)} - \bm a^{(k)}  \right)
\end{equation}
and with \eqref{stackcond_pa} and the definition\footnote{Note, that ${\bm {\tilde R}^{aa^{(k)}}}$ represents local properties. The suffix (k) will be dropped in the following.}
${\bm {\tilde R}^{aa^{(k)}}}  = \left( {\bm Q^{aa}}^{(k)} \right)^{-1}$ this gives:
\[
\bm {\bar p}^a  = \averg{ {\bm {\tilde R}^{aa}}  \left(   {\bm q^a}^{(1)} - \bm a^{(k)}  \right)} 
 = \averg{ \bm {\tilde R}^{aa}} {\bm q^a}^{(1)}   -  \averg{ \bm {\tilde R}^{aa}  \bm a^{(k)}  } 
\]
\[
{\bm q^a}^{(1)}   =  \averg{ \bm {\tilde R}^{aa}}^{-1} \bm {\bar p}^a +   \averg{ \bm {\tilde R}^{aa}}^{-1} \averg{ \bm {\tilde R}^{aa}  \bm a^{(k)}  } 
\]
We define $\widehat{\bm Q}^{aa} =  \averg{ \bm {\tilde R}^{aa}}^{-1}$ and write
\[
{\bm q^a}^{(1)}   = \widehat{\bm Q}^{aa} \bm {\bar p}^a +  \widehat{\bm Q}^{aa}  \averg{ \bm {\tilde R}^{aa}  \bm a^{(k)}  }, 
\]
which we insert into \eqref{A5}:
\begin{equation} 
 {\bm p^a}^{(k)}
= {\bm {\tilde R}^{aa}}  \left(   \widehat{\bm Q}^{aa} \bm {\bar p}^a +  \widehat{\bm Q}^{aa}  \averg{ \bm {\tilde R}^{aa}  \bm a^{(k)}  } - \bm a^{(k)}  \right)
\end{equation}
After rearranging this, we find
\begin{align} 
 {\bm p^a}^{(k)}
=\: &\: {\bm {\tilde R}^{aa}}   \widehat{\bm Q}^{aa} \bm {\bar p}^a 
+ {\bm {\tilde R}^{aa}}  \widehat{\bm Q}^{aa}  
\averg{  \bm {\tilde R}^{aa} {\bm Q^{ab}}^{(k)}} {\bm {\bar p}^b}
- {\bm {\tilde R}^{aa}} {\bm Q^{ab}}^{(k)} {\bm {\bar p}^b} \nonumber \\
&+ {\bm {\tilde R}^{aa}}  \widehat{\bm Q}^{aa}  
\averg{  \bm {\tilde R}^{aa}  {{\bm q^{\rm s}}^a}^{(k)} } 
- {\bm {\tilde R}^{aa}} {{\bm q^{\rm s}}^a}^{(k)}  
\end{align}
By comparison with \eqref{A2} we find the following expressions for $\bm A^{aa}$, $\bm A^{ab}$ and ${\bm p^{a}}^A$
{\setcounter{eqnstop}{\arabic{equation}}
\addtocounter{eqnstop}{1} % bisher steht da der letzte Wert drin!
\setcounter{equation}{0}
\renewcommand{\theequation}{\Alph{section}.\arabic{eqnstop}.\alph{equation}}
\begin{align} 
\bm A^{aa} & = \;\; {\bm {\tilde R}^{aa}}   \widehat{\bm Q}^{aa} \label{A9a}\\
\bm A^{ab} & = \;\; {\bm {\tilde R}^{aa}}  \widehat{\bm Q}^{aa}  
\averg{  \bm {\tilde R}^{aa} {\bm Q^{ab}}^{(k)}} - {\bm {\tilde R}^{aa}} {\bm Q^{ab}}^{(k)} \label{A9b} \\
{\bm p^{a}}^A & = \;\; {\bm {\tilde R}^{aa}}  \widehat{\bm Q}^{aa}  
\averg{  \bm {\tilde R}^{aa}  {{\bm q^{\rm s}}^a}^{(k)} } 
- {\bm {\tilde R}^{aa}} {{\bm q^{\rm s}}^a}^{(k)}  \label{A9c}
\end{align}
\setcounter{equation}{\arabic{eqnstop}}
} 
Equation \eqref{A9a} is equivalent to \eqref{sum_arbsol_b},  equation \eqref{A9b} is equivalent to \eqref{sum_arbsol_c} and equation \eqref{A9c} is equivalent to \eqref{sum_arbsol_e}.

Using this notation we can also express the effective properties in a symbolic way:
\begin{align*}
\bm Q^\ast = \left(  \begin{array}{cc}  
             {\bm Q^{aa}}^\ast & {\bm Q^{ab}}^\ast   \\
             {\bm Q^{ba}}^\ast & {\bm Q^{bb}}^\ast
 \end{array} \right)
& = \averg{\left(  \begin{array}{cc}  
             {\bm Q^{aa}}^{(k)} & {\bm Q^{ab}}^{(k)}   \\
             {\bm Q^{ba}}^{(k)} & {\bm Q^{bb}}^{(k)}
 \end{array} \right)
 \left(  \begin{array}{cc}  
             {\bm A^{aa}}^{(k)} & {\bm A^{ab}}^{(k)}   \\
             {\bm A^{ba}}^{(k)} & {\bm A^{bb}}^{(k)}
 \end{array} \right)
 }\\[1ex]
% = & \averg{\left( \begin{array}{cc}  
%   {\bm Q^{aa}}^{(k)} {\bm A^{aa}}^{(k)} \pl\pl & {\bm Q^{aa}}^{(k)} {\bm A^{aa}}^{(k)} + {\bm Q^{ab}}^{(k)}  \\
%   {\bm Q^{ba}}^{(k)} {\bm A^{aa}}^{(k)} \pl\pl & {\bm Q^{ba}}^{(k)} {\bm A^{ab}}^{(k)} + {\bm Q^{bb}}^{(k)} 
% \end{array} \right)
%}\\
{\bm Q^{aa}}^\ast &= \averg{ \bm Q^{aa} \bm A^{aa} } 
= \averg{ \bm Q^{aa} \bm {\tilde R}^{aa} }  \widehat{\bm Q}^{aa}
=  \widehat{\bm Q}^{aa} 
\\[1ex]
{\bm Q^{ab}}^\ast &= \averg{ \bm Q^{aa} \bm A^{ab} } + \averg{ \bm Q^{ab} } 
= \widehat{\bm Q}^{aa} \averg{\bm {\tilde R}^{aa} \bm Q^{ab} }  
\\[1ex]
{\bm Q^{ba}}^\ast &= \averg{ \bm Q^{ba}  \bm A^{aa} } 
= \averg{ \bm Q^{ba} \bm {\tilde R}^{aa} } \widehat{\bm Q}^{aa}  
= \left( {\bm Q^{ab} }\right)^t \\[1ex]
{\bm Q^{bb}}^\ast &= \averg{ \bm Q^{ba} \bm A^{ab}  + \bm Q^{bb}  } \\
&= \averg{ \bm Q^{ba}  \bm {\tilde R}^{aa} } \widehat{\bm Q}^{aa} \averg{ \bm {\tilde R}^{aa}  \bm Q^{ab}  } - \averg{ \bm Q^{ba}  \bm {\tilde R}^{aa}   \bm Q^{ab}  } + \averg{\bm Q^{bb} }
\\
\intertext{And the effective spontaneous strain and polarization are:}
\bm q^{\rm s \ast} = \left(\begin{array}{c} {\bm q^{\rm s\ast}}^a \\  {\bm q^{\rm s\ast}}^b  \end{array} \right)
&= \averg{\left(  \begin{array}{cc}  
             {\bm Q^{aa}}^{(k)} & {\bm Q^{ab}}^{(k)}   \\
             {\bm Q^{ba}}^{(k)} & {\bm Q^{bb}}^{(k)}
\end{array} \right)
\left(  \begin{array}{c}  
              {\bm p^A}^{a\, (k)}   \\
              {\bm p^A}^{b\, (k)} 
\end{array} \right)} 
+ \averg{\left(\begin{array}{c} {\bm q^{\rm s}}^a \\  {\bm q^{\rm s}}^b  \end{array} \right)} 
\\[1ex]
{\bm q^{\rm s\ast}}^a & = \averg{ \bm Q^{aa} {\bm p^A}^{a}} + \averg{{\bm q^{\rm s}}^a} 
= \widehat{\bm Q}^{aa} \averg{ \bm {\tilde R}^{aa}  {\bm q^{\rm s}}^{a} }  
\\[1ex]  
{\bm q^{\rm s\ast}}^b & = \averg{{\bm Q^{ba}}  \bm {\tilde R}^{aa}}  \widehat{\bm Q}^{aa}
\averg{ \bm {\tilde R}^{aa}  {\bm q^{\rm s}}^a} 
- \averg{{\bm Q^{ba}}  \bm {\tilde R}^{aa}  {\bm q^{\rm s}}^a} + \averg{ {\bm q^{\rm s}}^b  }.
\end{align*}

It seems not to be useful to give explicit equations for the various components of this general solution, as the expressions are quite lengthy. We have not assumed any symmetry relation for the crystal and its property tensors and will therefore not expect to find any particular symmetry changes in this results.  
It should be noted that the matrices $ {\bm Q^{aa}}^{(k)}$, $ {\bm Q^{ab}}^{(k)}$ and $\bm {\tilde R}^{aa}$ are constants for the various phases $k$. On the contrary, $\widehat{\bm Q}^{aa}$ is a over-all property which depends on the volume fractions $\xi^{(k)}$. Finally, the interaction matrices ${\bm A^{aa}}^{(k)}$ and ${\bm A^{ab}}^{(k)}$ hold for a particular phase $k$, but depend on the volume fraction $\xi^{(p)}$ of all phases and are not particular properties of the phase $k$ alone.

We can proof the general solution by showing that the micro-macro relations are fulfilled for all components:
\begin{align*}
\bm {\bar p} &= \averg{\bm p} = \averg{\bm A} \bm {\bar p} + \averg{\bm p^A}
\end{align*}
which is true, if $\averg{\bm A} = \bm I $ and $\averg{\bm p^A} = \bm 0$. This can be shown easily.
Additionally, one can show, that $\bm q^a$ are homogeneous, i.e. they depend on macroscopic quantities only:
\begin{align*}
 \bm q^a &= \bm Q^{aa}\bm p^a + \bm Q^{ab} \bm p^b + {\bm q^s}^a 
= \widehat{\bm Q}^{aa}  \left( {\bm{\bar p}^a} + \averg{\tilde {\bm R} \bm Q^{ab}} {\bm{\bar p}^b}
+ \averg{\tilde {\bm R} {\bm q^{\rm s}}^a }     \right)
\end{align*}

{\it Homogeneous material properties}

If $\bm Q$ is homogeneous, i.e. all lamellae have the same linear properties, then follows:
\begin{align*}
\widehat{\bm Q}^{aa} & = \bm Q^{aa}
\\
\bm A^{aa} & = \bm {\tilde R}^{aa}  \, \widehat{\bm Q}^{aa}  = \bm I^{aa}
\\
\bm A^{ab} & = \bm 0,\\
\intertext{where $\bm I^{aa}$ is a $4\times 4$ identity matrix, 
which shows, as expected, that $\bm A = \bm I$, with $\bm I$ beeing a $9\times 9$ identity matrix. Then}
\bm Q^\ast &= \averg{\bm Q  \bm A}  = \bm Q.
\\ 
\intertext{If $\bm Q$ is homogeneous, but the spontaneous strain and polarization $\bm q^{\rm s}$ are not homogenoeus, then}
{\bm p^{a}}^A & =  {\bm {\tilde R}^{aa}}  \left(
\averg{ {{\bm q^{\rm s}}^a}^{(k)}} -  {{\bm q^{\rm s}}^a}^{(k)} \right)
\end{align*}
which says, that the local stress and electric field are proportional to the difference between the average spontaneous strain and polarization and the local one. We find that for homogenous linear properties the effective spontaneous strain and polarization are equal to their simple volume average:
\begin{align} 
{\bm q^{\rm s\ast}} & = \averg{ {\bm q^{\rm s}} }  
\label{average_qs_appendix} 
\end{align}

{\it Partial homogeneity}

It is interesting, and for the case of ferroelectric domain structures important to note, that considerable simplifications of the general equations for the effective properties may be found, even if not all components of $\bm Q$ are homogeneous. If only the $\bm Q^{aa}$, and therefore $\tilde{\bm R}^{aa^{(k)}}$, are homogeneous, then $\widehat{\bm Q}^{aa}  {\bm {\tilde R}^{aa}}   = \bm I^{aa}$ and we get  following relationships:
\begin{align}
\bm A^{aa} & = \bm {\tilde R}^{aa}  \, \widehat{\bm Q}^{aa}  = \bm I    \label{A_aa_twin}\\
\bm A^{ab} & =\bm {\tilde R}^{aa}  \left(  \averg{ {\bm Q^{ab}}^{(k)}} -{\bm Q^{ab}}^{(k)}  \right) \label{A_ab_twin} \\
{\bm p^{a}}^A & = \;\; {\bm {\tilde R}^{aa}}  
\left(  \averg{ {{\bm q^{\rm s}}^a}^{(k)} } -  {{\bm q^{\rm s}}^a}^{(k)}  \right). \label{p_Aa_twin} \\
\intertext{and for the effective properties}
{\bm Q^\ast}^{aa} & = \averg{\bm Q^{aa}} = \bm Q^{aa}
\\[1ex]
{\bm Q^\ast}^{ab} & = \averg{\bm Q^{ab}} 
\\[1ex]
{\bm Q^\ast}^{ba} & = \averg{\bm Q^{ba}} = \left(  {\bm Q^\ast}^{ab} \right)^t 
\\[1ex]
{\bm Q^\ast}^{bb} & = \averg{ \bm Q^{ba}}  \bm {\tilde R}^{aa} \averg{  \bm Q^{ab}  } - \averg{ \bm Q^{ba}  \bm {\tilde R}^{aa}   \bm Q^{ab}  } + \averg{\bm Q^{bb} }
\\[1ex]
{\bm q^{\rm s\ast}}^a & = \averg{ {\bm q^{\rm s}}^a }  
\\[1ex]
{\bm q^{\rm s\ast}}^b & = \averg{\bm Q^{ba}}  \bm {\tilde R}^{aa}  \averg{ {\bm q^{\rm s}}^a} 
- \averg{{\bm Q^{ba}}  \bm {\tilde R}^{aa}  {\bm q^{\rm s}}^a} + \averg{ {\bm q^{\rm s}}^b  }.\\
\intertext{If additionally the components ${\bm q^s}^a$ are homogeneous, then the latter equation simplies to}
{\bm q^{\rm s\ast}}^b & = \averg{ {\bm q^{\rm s}}^b  }
\end{align}

For further discussions it is useful to split the effective linear properties into the simple average $\averg{\bm Q} $ and an additional contribution $\bs \Lambda$. It is then convenient to express the interaction matrix by $\bm A = \bm I + \bm A'$: 
\begin{align*}
\bm A' &= \bm I +   \left(  \begin{array}{cc}  
             \bm 0 & {\bm A^{ab}}   \\
             \bm 0 &  \bm 0
 \end{array} \right).
\intertext{Then}
\bm Q^\ast &= \averg{\bm Q \, \bm A} =  \averg{\bm Q \, \left(\bm I + \bm A' \right)} = \averg{\bm Q } + \averg{\bm Q \bm A' } \\[1ex]
\bm \Lambda &= \averg{\bm Q \bm A' } = \bm Q^\ast -  \averg{\bm Q } = \left(  \begin{array}{cc}  
             \bm 0 & \bm 0  \\
             \bm 0 & \bm \Lambda^{bb}
 \end{array} \right)\\[1ex]
\bm \Lambda^{bb} &=  \averg{ \bm Q^{ba}}  \bm {\tilde R}^{aa} \averg{  \bm Q^{ab}  } - \averg{ \bm Q^{ba}  \bm {\tilde R}^{aa}   \bm Q^{ab}  }
\end{align*}

\newpage

{

%\bibliography{../bib/piezo,../bib/elast,../bib/reports,../bib/plast,../bib/multiferro,../bib/buecher}

\begin{thebibliography}{33}
\expandafter\ifx\csname natexlab\endcsname\relax\def\natexlab#1{#1}\fi
\expandafter\ifx\csname url\endcsname\relax
  \def\url#1{\texttt{#1}}\fi
\expandafter\ifx\csname urlprefix\endcsname\relax\def\urlprefix{URL }\fi

\bibitem[{Aksakaya and Farnell(1988)}]{akcakaya.88}
Aksakaya, E., Farnell, G.~W., 1988. Effective elastic and piezoelectric
  constants of superlattices. J. Appl. Phys. 64~(9), 4469--4473.

\bibitem[{Arlt(1990)}]{arlt.90}
Arlt, G., 1990. Twinning in ferroelectric and ferroelastic ceramics: Stress
  relief. J. mater. sci. 25, 2655--2666.

\bibitem[{Arlt and Sasko(1980)}]{arlt.80}
Arlt, G., Sasko, P., 1980. Domain configuration and equilibrium size of domains
  in {$\rm BaTiO_3$} ceramics. J. Appl. Phys. 51, 4956--4960.

\bibitem[{Avellaneda and Harsh\'{e}(1994)}]{avellaneda.94}
Avellaneda, M., Harsh\'{e}, G., 1994. Magnetoelectric effect in
  piezoelectric/magnetostrictive multilayer (2-2) composites. J. Intell. Mat.
  Struct. 5~(4), 501-- 513.

\bibitem[{Bednarcyk(2003)}]{bednarcyk.03}
Bednarcyk, B.~A., 2003. An inelastic micro/macro theory for hybrid smart/metal
  composites. Composites B 34~(2), 175--197.

\bibitem[{Bhattacharya and Ravichandran(2003)}]{bhattacharya.03}
Bhattacharya, K., Ravichandran, G., 2003. Ferroelectric perovskites for
  electromechanical actuation. Acta Materialia 51, 5941--5960.

\bibitem[{Cao et~al.(1993)Cao, Zhang, and Cross}]{cao.93b}
Cao, W.~W., Zhang, Q.~M., Cross, L.~E., 1993. Theoretical study on the static
  performance of piezoelectric ceramic polymer composites with 2-2
  connectivity. IEEE Trans. Ultrasonics, Ferroelelectrics and Frequency Control
  40~(2), 103--109.

\bibitem[{Davis et~al.(2005)Davis, Damjanovic, Hayem, and Setter}]{davis.05}
Davis, M., Damjanovic, D., Hayem, D., Setter, N., 2005. Domain engineering of
  the transverse piezoelectric coefficient in perovskites ferroelectrics. J.
  Appl. Phys. 98, 014102.

\bibitem[{Erhart and Cao(1999)}]{erhart.99}
Erhart, J., Cao, W., 1999. Effective material properties in twinned
  ferroelectric crystals. J. Appl. Phys. 86, 1073--1081.

\bibitem[{Erhart and Cao(2001)}]{erhart.01}
Erhart, J., Cao, W., 2001. Effective symmetry and physical properties in
  twinned perowskite ferroelectric single crystals. J. Mater. Res. 16,
  570--577.

\bibitem[{Gibiansky and Torquato(1999)}]{gibiansky.99}
Gibiansky, L.~V., Torquato, S., 1999. Matrix laminate composites: Realzable
  approximations for the effective moduli of piezoelectric dispersions. J.
  Mater. Res. 14~(1), 49--63.

\bibitem[{Janas and Safari(1995)}]{janas.95}
Janas, V.~F., Safari, A., 1995. Overview of fine-scale piezoelectric
  ceramic/polymer composite processing. J. Am. Ceram. Soc. 78~(11), 2945--55.

\bibitem[{Koh et~al.(2005)Koh, Yoon, Lee, and Kim}]{koh.05}
Koh, Y.-H., Yoon, C.-B., Lee, S.-M., Kim, H.-E., 2005. Thermoplastic green
  machining for the fabrication of a piezoelectric ceramic/polymer composite
  with 2-2 connectivity. J. Am. Ceram. Soc. 88~(4), 1060--1063.

\bibitem[{Kreher(1998)}]{report.MW282710}
Kreher, W., Juli 1998. {Effektive Konstanten Schichtverbund}. Tech. Rep.
  MW282.710, TU Dresden, Inst. f. Werkstoffwissenschaft, Professur
  Materialwissenschaft und Nanotechnik.

\bibitem[{Li and Liu(2004)}]{li.04}
Li, J.~Y., Liu, D., 2004. On ferroelectric crystals with engineered domain
  configurations. J. Mech. Phys. Solids 52, 1719--1742.

\bibitem[{Liu and Li(2003)}]{liu-li.03}
Liu, D., Li, J.~Y., 2003. The enhanced and optimal piezoelectric coefficients
  in single crystalline barium titanate with engineered domain configurations.
  Appl. Phys. Letters 83, 1193--1195.

\bibitem[{Liu and Li(2004)}]{liu-li.04}
Liu, D., Li, J.~Y., 2004. Domain engineered {$\rm
  Pb(Mn_{1/3}Nb_{2/3})O_3-PbTiO_3$} crystals: Enhanced piezoelectricity and
  optimal domain configurations. Appl. Phys. Letters 84~(19), 3930--3932.

\bibitem[{Lous et~al.(2000)Lous, Cornejo, McNulty, Safari, and
  Danforth}]{lous.00}
Lous, G.~M., Cornejo, I.~A., McNulty, T.~F., Safari, A., Danforth, S.~C., 2000.
  Fabrication of piezoelectric ceramic/polymer composite transducers using
  fused deposition of ceramics. J. Am. Ceram. Soc. 83~(1), 124--128.

\bibitem[{Milton(2002)}]{milton.02}
Milton, G.~W., 2002. The Theory of Composites. Cambridge Univ. Press,
  Cambridge, UK.

\bibitem[{Nan et~al.(2005)Nan, Cai, Shi, Zhai, Liu, and Lin}]{nan.05}
Nan, C., Cai, N., Shi, Z., Zhai, J., Liu, G., Lin, Y., 2005. Large
  magnetoelectric response in multiferroic polymer-based composites. Phys. Rev.
  B 71, 014102.

\bibitem[{Newnham et~al.(1978)Newnham, Skinner, and Cross}]{newnham.78}
Newnham, R.~E., Skinner, D.~P., Cross, L.~E., 1978. Connectivity and
  piezoelectric-pyroelectric composites. Mat. Res. Bul. 13, 525--536.

\bibitem[{Nye(1964)}]{nye}
Nye, J.~F., 1964. Physical Properties of Crystals. Clarendon Press, Oxford.

\bibitem[{Park et~al.(1999)Park, Wada, Cross, and Shrout}]{park.99}
Park, S.-E., Wada, S., Cross, L.~E., Shrout, T.~R., 1999. Crystallographically
  engineered batio$_3$ single crystals for high-performance piezoelectrics. J.
  Appl. Phys. 86, 2746--2750.

\bibitem[{R{\"o}del and Kreher(2003)}]{roedel.03}
R{\"o}del, J., Kreher, W.~S., 2003. Modelling linear and nonlinear behavior of
  polycrystalline ferroelectric ceramics. J. Europ. Ceramic Soc. 23, 2297 --
  2306.

\bibitem[{Shu and Bhattacharya(2001)}]{shu.01}
Shu, Y.~C., Bhattacharya, K., 2001. Domain patterns and macroscopic behaviour
  of ferroelectric materials. Phil. Mag. B 81, 2021--2054.

\bibitem[{Turik(1970)}]{turik.70engl}
Turik, A.~V., 1970. Elastic, piezoelectric and dielectric properties of single
  crystals of $\rm batio_3$ with laminar domain structures. Soviet Physics -
  Solid State 12, 688--693.

\bibitem[{Turik and Bondarenko(1974)}]{turik.74b}
Turik, A.~V., Bondarenko, E.~I., 1974. Effect of domain-strructure on physikal
  properties of ferroelectrics. Ferroelectrics 7~(1-4), 303--305.

\bibitem[{Wada et~al.(1999)Wada, Suzuki, Noma, Suzuki, Osada, Kakihana, Park,
  Cross, and Shrout}]{wada.99}
Wada, S., Suzuki, S., Noma, T., Suzuki, T., Osada, M., Kakihana, M., Park,
  S.-E., Cross, L.~E., Shrout, T.~S., 1999. Enhanced piezoelectric properties
  of barium titanate single crystals with engineered domain configurations.
  Jpn. J. Appl. Phys. 38, 5505--5511.

\bibitem[{Wada et~al.(2005)Wada, YXako, Kakemoto, and Tsurumi}]{wada.05}
Wada, S., YXako, K., Kakemoto, H., Tsurumi, T., 2005. Enhanced piezoelectric
  properties of barium titanate single crystals with different
  engineered-domain sizes. J. Appl. Phys. 98, 014109.

\bibitem[{Yin and Cao(2002)}]{yin.02}
Yin, J., Cao, W., 2002. Effective macroscopic symmetries and materials
  properties of multidomain {$\rm 0.955 Pb(Zn_{1/3}Nb_{2/3})O_3 - 0.045
  PbTiO_3$} single crystals. J. Appl. Phys. 92~(1), 444--448.

\bibitem[{Zgonik et~al.(1994)Zgonik, Bernasconi, Duelli, Schlesser, G\"unter,
  Garrett, Rytz, Zhu, and Wu}]{zgonik.94}
Zgonik, M., Bernasconi, P., Duelli, M., Schlesser, R., G\"unter, P., Garrett,
  M.~H., Rytz, D., Zhu, Y., Wu, X., 1994. Dielectric, elastic, piezoelectric,
  electro-optic and elasto-optic tensors of {BaTiO$_3$} crystals. Phys. Rev. B
  50, 5941--49.

\bibitem[{Zhang et~al.(1994)Zhang, Cao, Zhao, and Cross}]{zhang.94b}
Zhang, Q.~M., Cao, W.~W., Zhao, J., Cross, L.~E., 1994. Piezoelectric
  performance of piezoceramic-polymer composites with 2-2 connectivity - a
  combined theoretical and experimental-study. IEEE Trans. Ultrasonics,
  Ferroelelectrics and Frequency Control 41~(4), 556--564.

\bibitem[{Zhang et~al.(2001)Zhang, Jiang, and Cao}]{zhang.01}
Zhang, R., Jiang, B., Cao, W., 2001. Elastic, piezoelectric and dielectric
  properties of multidomain {$\rm 0.67 Pb (Mg_{1/3}Nb_{2/3}O_3) - 0.33
  PbTiO_3$} single crystals. J. Appl. Phys. 90, 3471.

\end{thebibliography}
}

\newpage

\listoftables

\listoffigures

%%%%%%%%%%%%%%%%%%%%%%%%%%%%%%%%%%%%%%%%%%%%%%%%%%%%%%%%%%%%%%%%%%%%%%%%%%%%%

\end{document}